\documentclass[runningheads]{svmult}

\usepackage{makeidx}   
\usepackage{graphicx}  
\usepackage{subeqnar}  
\usepackage{multicol}  
\usepackage{physprbb}  
\makeindex             

\def\kvec{ {\bf k}}
\def\Qvec{ {\bf Q}}

\def\kktil{\widetilde {k}}

\def\qtil{\widetilde {q}}

\def\0til{\bar {0}}

\def\dm{\frac{1}{2}}

\def\sumslashD{\mathop{\sum \kern-1.4em -\kern 0.5em}}
\def\sumslash{\mathop{\sum \kern-1.2em -\kern 0.5em}}
\def\intslash{\mathop{\int \kern-0.9em -\kern 0.5em}}
\def\intslashD{\mathop{\int \kern-1.1em -\kern 0.5em}}

\begin{document}
\title*{Electronic phases of low dimensional conductors }

%
%
%
%
%
\author{C. Bourbonnais}
\authorrunning{C. Bourbonnais}
%
%
\institute{Centre de Recherche sur les Propri\'et\'es \'Electroniques  de
Mat\'eriaux Avanc\'ees, D\'epartement de Physique Universit\'e de Sherbrooke,
Sherbrooke, Qu\'ebec, Canada J1K 2R1 }

\maketitle              

\begin{abstract}
       We briefly review the physics of electronic phases in low dimensional conductors. We begin by introducing
the properties of  the one-dimensional electron gas model using  bosonization and   renormalization group
methods.We then tackle the influence of interchain coupling and go through the different instabilities of
the electron system to the formation of higher dimensional states.  The connection  with observations
made in quasi-one-dimensional organic and inorganic conductors is discussed.   
\end{abstract}

\section{Introduction}
There is a {\it consensus generalis} about the  impact of   reducing   spatial dimension in systems of interacting
electrons: correlation effects  are magnified and  range of  electronic behaviors expanded.  This is
well  exemplified in one spatial dimension where  low energy electronic excitations turn out to be 
entirely collective in character so that a description in terms of Fermi liquid quasi-particles,  which proved to
be so successful in  isotropic systems,  becomes   simply inapplicable. 
 
Organic conductors belong to  a class of crystals  for which the planar conformation of the 
molecular constituents combined with their packing as weakly coupled  chains
in the solid state yield close
realizations of one-dimensional solids.  Quasi-one
dimensional electronic structures  are also   found in  inorganic materials such as the 
molybdenum bronzes, the chalcogenides and ladder 
cuprates,\footnote{See the reviews of D. J\'erome and T. M. Rice  on ladder systems in this volume. Luttinger
liquid behavior from edge states of two-dimensional electron gas in quantum well structures is discussed by C.
Glattli in this volume.} as a result of  their peculiar molecular or atomic  arrangements.      In all these
quasi-one-dimensional  crystals, we are thus faced with a twofold difficulty  which combines the objective of
determining the temperature range where  one-dimensional physics applies with the one of finding the origin of
low temperature higher dimensional   states. 
   In this review, we will be mainly concerned with these two closely bound issues that are at the heart of
the  description of  the rich   phase diagram of compounds like the Bechgaard salt series
((TMTSF)$_2$X)  and their sulfur analogs, the Fabre salt series ((TMTTF)$_2$X, where X = PF$_6$, AsF$_6$ ... is  
a monovalent anion). 

In the first part of this review, we  shall briefly  outline  the non Fermi liquid properties of 
the one-dimensional electron gas using bosonization and renormalization group methods. We  then consider
the problem  of instabilities of one-dimensional electronic states  by introducing the coupling between
chains. We discuss subsequently how the concepts of low-dimensional physics prove   relevant when one
tries to construct a coherent picture of electronic states that are actually found in quasi-one-dimensional
organic compounds. We close this review with a brief discussion on the importance of one-dimensional physics in
  inorganic metals  like the molybdenum bronzes.

\section{Interacting electrons in one dimension}

\subsection{The Tomanaga-Luttinger model}
\label{TL}
 When one tries to understand  the origin of non-Fermi-liquid behavior in one dimension,\footnote{We refer to the
excellent reviews of J. Voit \cite{Voit95}, H. J. Schulz \cite{Schulz94} and V. J. Emery \cite{Emery79} for a
more exhaustive discussion  of the one-dimensional gas model using bosonization method. }  it is instructive to
look first  at the possible elementary excitations such a low dimensional system can sustain. Consider the
transfer of an electron  in the final state of   wave vector 
$k+q$ above a filled Fermi sea. Together with the hole left at
$k$, both particles form an  electron-hole elementary excitation of energy  
$\omega(q)=\epsilon(k+q)
-\epsilon(k)$ ($\hbar=1$). In one 
dimension, the available phase space below $\omega(q)$ for the 
  decay of such   excitations shrinks to  zero at low energy, namely where the spectrum
\begin{equation}
\epsilon(k)\to \epsilon_p(k)= v_F(pk-k_F)
\label{Luttinger}
\end{equation}
can be considered as {\it linear}. A linear spectrum with an unbounded interval of $k$ values for each  branch   
$p=\pm$ (refering to right ($+$) and  left ($-$) moving electrons)   defines the spectrum of the Luttinger
model.\cite{Luttinger63} In these conditions,  electron-hole excitations acquire  a high  degree of degeneracy 
 $\sim qL/2\pi$. It follows that stable charge and spin-density excitations can be formed from the   
superpositions of excitations  
\begin{eqnarray}
\rho_\pm(q) =   {1\over \sqrt{2}}\sum_{\alpha,k}\,
a^\dagger_{\pm,k+q,\alpha}a_{\pm,k,\alpha} 
\end{eqnarray}
for the charge and 
\begin{eqnarray}
\sigma_\pm(q) =    {1\over \sqrt{2}}\sum_{\alpha,k}\alpha\,
a^\dagger_{\pm,k+q,\alpha}a_{\pm,k,\alpha}
\end{eqnarray}
 for the  spin. This
connotes that a free -- bosonic --
Hamiltonian may exist for the description of collective charge and spin modes. 
Actually  for a Luttinger spectrum with  all negative energy states  filled,
\cite{Mattis65} these
composite  objects obey  the commutation rules  
\begin{eqnarray}
[\rho_p(q),\rho_{p'}(-q)]= && p\delta_{pp'}\,q {L\over 2\pi}\cr
[\sigma_p(q),\sigma_{p'}(-q)]= && p \delta_{pp'}\,q {L\over 2\pi},
\end{eqnarray}
which are compatible with bosons. Another key feature is the commutation relation
between the  one-electron
Hamiltonian
\begin{equation}
H_0 = \sum_{k,p,\alpha} \epsilon_p(k) a^\dagger_{p,k,\alpha}a_{p,k,\alpha}
\end{equation}
and the spin and charge-density operators:
\begin{eqnarray}
[H_0,\rho_p(q)]=  && pv_Fq\,\rho_p(q)\cr
[H_0,\sigma_p(q)]=  && pv_Fq\,\sigma_p(q),
\end{eqnarray}
which reminds the algebra of operators for  the harmonic oscillator. A 
fermion to boson correspondance for the
excitations can  thus be established in which $H_0$ can be   written as
\begin{eqnarray}
H_0=   \sum_{p,q} \, pv_Fq \big(\bar{\rho}_p(q)\bar{\rho}_p(-q) +
\bar{\sigma}_p(q)\bar{\sigma}_p(-q)\big),
\end{eqnarray}
which is quadratic in  the charge $\bar{\rho}_p\equiv 
(2\pi/Lq)^\dm\rho_p $ and spin
$\bar{\sigma}_p\equiv (2\pi/Lq)^\dm\sigma_p$ operators. All the excited states of the fermion system can then be
described in terms of bosonic variables.

In the Tomanaga-Luttinger model,\cite{Tomonaga50,Luttinger63} electrons interact through the exchange
of small momentum transfer. This allows
us to define two scattering processes usually denoted $g_2$
and
$g_4$ couplings with respect    to the  Fermi
points $\pm k_F$ (here taken for simplicity as $q-$independent interactions 
\cite{Mattis65,Schulz83}). The total
Hamiltonian becomes
$H_{TL}= H_0 +H_I
$, in which the interacting
part $H_I$ can also be expressed in terms of density operators:
\begin{eqnarray}
H_I  &=&   {1\over L} \sum_{\alpha_{1,2}}\sum_{k_1,k_2,q} g_2
\,a_{+,k_1+q,\alpha_1}^\dagger
a_{-,k_2-q,\alpha_2}^\dagger a_{-k_2,\alpha_2}a_{+,k_1,\alpha_1}\cr
    &+ &   {1\over L} \sum_{\alpha,p}\sum_{k_1,k_2,q} g_4\,
a_{p,k_1+q,\alpha }^\dagger
a_{p,k_2-q,-\alpha}^\dagger a_{pk_2,-\alpha}a_{p,k_1,\alpha}\cr
   & = &     2 g_2\sum_{p,q} v_Fq \bar{\rho}_p(q) \bar{\rho}_{-p}(-q) +  g_4
\sum_{p,q} v_Fq\big(\bar{\rho}_p(q) \bar{\rho}_{p}(-q) -
\bar{\sigma}_p(q)
\bar{\sigma}_{p}(-q)\big).
\end{eqnarray}
The TL Hamiltonoian   is therefore still quadratic in terms of the bosonic 
operators. Owing to
the  presence of the $g_2$
term which couples the density on different branches,   a 
transformation is
needed to diagonalize  the  Hamiltonian. We thus have
\begin{eqnarray}
H_{TL} &= &     H_\sigma + H_\rho\cr
    &= & \sum_{p,q} \omega_\sigma(q) \, b^\dagger_{\sigma,q} b_{\sigma,q} +
\omega_\rho(q) b^\dagger_{\rho,q}
b_{\rho,q},
\end{eqnarray}
in which the new operators $b^{(\dagger)}_{\sigma}$ for spin and
$b^{(\dagger)}_{\rho}$ for charge obey
boson commutation rules. The spin and charge  spectra of collective excitations have a
Debye   form \hbox{$\omega_{\nu}(q)=
u_\nu \mid q\mid $} ($\nu=\sigma,\rho)
$, where the velocities are
\begin{equation}
u_\sigma= v_F~\Big(1- {g_4\over 2\pi v_F} \Big)
\label{veloTLspin}
\end{equation}
for the spin   and
\begin{equation}
u_\rho= v_\rho\left(1  - \left({g_2\over\pi v_\rho}\right)^2\right)^{1/2}
\label{veloTLcharge}
\end{equation}
for the charge, with $v_\rho = v_F(1+g_4/2\pi v_F)$.
A key property of the above Hamiltonian is  the commutation relation 
$[H_\sigma,H_\rho]=0 $, which yields the decoupling between  $H_\sigma$ and $H_\rho$ that is  termed  {\it
separation between  spin and charge degrees of freedom}. The corresponding split-off of acoustic excitations will
have a profound influence  on the properties of
the system which becomes  a Luttinger liquid.\cite{Haldane81} As regards thermodynamics, the free energy 
will consist  of two separate contributions, which yields the property of additivity for
the 
   specific heat (per unit of length) $C=C_\sigma + C_\rho$.    
Here $C_{\nu=\sigma,\rho} = \pi k_B^2T/(3u_\nu)$ is the  linear  
temperature dependent  specific heat of each branch of acoustic
excitations.

\subsection{Phase variables description}
   Collective excitations in a Luttinger liquid
are reminiscent of those   of a
vibrating string. This relation can be further sharpened if one introduces the   
 pair of conjugate phase variables
\begin{eqnarray}
\phi_\nu(x) &= &  -i {\pi\over L}\sum_q {e^{-iqx}\over q} 
e^{-\alpha_0 \mid q\mid/2} [\nu_+(q)
+\nu_-(q)]\cr
\Pi_\nu(x) &= & i {1\over L}\sum_q   e^{-iqx}e^{-\alpha_0 \mid 
q\mid/2} [\nu_+(q)
-\nu_-(q)],
\end{eqnarray}
which satisfy commutation relation $[\phi_\nu(x_1),\Pi_{\nu'}(x_2)]= 
i \delta_{\nu,\nu'}\delta(x_1-x_2)$ in
  the  limit
$\alpha_0\to 0$ for the  short distance cut-off.

The phase variable
representation
     allows us to rewrite the  Tomonaga-Luttinger Hamiltonian in the harmonic form
\begin{equation}
H_{TL}= \sum_{\nu =\rho,\sigma}\dm \int  \left[ \pi u_\nu K_\nu \Pi_\nu^2 +
u_\nu (\pi K_\nu)^{-1}
\left({\partial
\phi_\nu \over \partial x}\right)^2 \right] dx,
\label{phaseH}
\end{equation}
where $K_\nu$  is the stiffness constants of  acoustic  excitations. 
The properties of the model are
then entirely  governed by the set of parameters $\{u_\nu,K_\nu\}$.
 These are functions of the
microscopic coupling constants.  Thus for a rotationally invariant system (spin independent interactions),
one has $K_\sigma=1$ and 
\begin{equation}
K_\rho=  \left(2\pi v_F + g_4 -2g_2\over 2\pi v_F + g_4 + 2g_2\right)^{1/2}.
\end{equation}

\subsection{Properties of the Luttinger liquid state}
\label{PropLL}

\subsubsection{One-particle}
Consider the Matsubara time-ordered single-particle Green's function 
$G_p(x,\tau)  =  -\,\langle
\,T_\tau \,\psi_{p,\alpha}(x,\tau)\psi_{p,\alpha}^\dagger(0,0)\, 
\rangle$, which is expressed as a
statistical average over   fermion fields. It can be    evaluated 
explicitly by using the harmonic phase
Hamiltonian (\ref{phaseH}),  with the aid  of the relation
between the Fermi and bosonic fields \cite{Mattis74,Luther74,Haldane80}:
\begin{eqnarray}
\psi_{p,\alpha}(x)  &=&  L^{-\dm} \sum_k a_{p,k,\alpha}\ e^{ikx}\cr
&\sim &\lim_{\alpha_0 \to 0} {e^{ipk_Fx}\over \sqrt{2\pi\alpha_0}} 
\exp\Bigl(-{i\over
\sqrt{2}}[p(\phi_\rho + \alpha \phi_\sigma) + (\theta_\rho + 
\alpha\theta_\sigma)]\Bigr),
\label{bosonization}
\end{eqnarray}
where $\theta_\nu(x)=\pi\int \Pi_\nu(x') dx'$. One finds
\begin{eqnarray}
G_p(x,\tau) =  {e^{ipk_Fx}\over   \alpha_0^{-\theta}} \prod_{\nu}
[\xi_\nu\sinh(x+iu_{\nu}\tau)/\xi_\nu]^{-{1\over
2}-\theta_\nu}[\xi_\nu\sinh(x-iu_{\nu}\tau)/\xi_\nu]^{-\theta_\nu}
\label{thermalG}
\end{eqnarray}
    where $\theta_\nu = {1\over 4} (K_\nu + 1/K_\nu -2)$. As
a correlation function of the
electron with itself, the Green's function  gives useful information  about
the spatial and time decay of
single-particle quantum coherence in the presence of collective 
oscillations of the Luttinger liquid. At equal
Matsubara time, which amounts to put $\tau=0$ in the above expression, we observe that $G_p$   
 depends on two
characteristic length scales
$\xi_\sigma= u_\sigma/\pi T$ and
$\xi_\rho= u_\rho/\pi T$, corresponding to the de Broglie quantum lengths for
spin and charge acoustic
excitations. Thus for     $ \alpha_0\ll x \ll \xi_\nu$, the
fermion coherence  decays according to the power law
\begin{equation}
G_p(x) \approx {e^{ipk_Fx}\over   \alpha_0^{-\theta}} {1\over x^{1+\theta}},
\label{G(x)}
\end{equation}
where the exponent $\theta=\theta_\sigma + \theta_\rho$ is called the
anomalous  dimension of the Green's
function. It is non-zero in   the presence of interaction (the 
canonical dimension of the
Green's function is unity in  a free electron gas). For non-zero 
interaction, the spatial decay of  
quasi-particle
    coherence is therefore faster. The existence of a
anomalous power law is also the mark of scaling (a situation  analogous to
 one of  correlations of the order
parameter  at the critical point), namely  the absence of particular length
scale of the fermion
coherence between
$\alpha_0$ and $\xi_\nu$.

For large distances  $x\gg \xi_\nu$,
we have
\begin{equation}
G_p(x) \propto e^{-x/\xi},
\end{equation}
indicating that thermal fluctuations lead to an exponential decay of 
coherence and  the
absence of scaling. The effective
coherence length  $\xi^{-1} = 1/\xi_\sigma + 1/\xi_\rho$
combines the spin and the charge
quantum lengths.

The absence of ordinary quasi-particle states in a Luttinger liquid will also
show up in the one-electron
spectral properties. These can be extracted from the Fourier transform of the retarded Green's function. The 
quantity of interest is the spectral
weight defined as the imaginary part of the Green's
function:
\begin{equation}
A_p(q,\omega)= -{1\over \pi} {\rm Im}G_p(pk_F+q,\omega),
\end{equation}
    which  gives the probability of having a single-particle state of wave
vector $pk_F +q$ with a energy $\omega$
measured from  the Fermi level. The spectral function  takes  on particular importance since  it can be probed
at
$q<0\,  (q>0)$  by photoemission
(inverse photoemission) experiments.  We will focus here on   spin
independent interactions for which $K_\sigma=1$ and 
$\theta_\sigma=0$. The presence of collective modes with
two different velocities in a Luttinger liquid has a pronounced 
influence on the  spectral function
in comparison to that of a  Fermi liquid.
In the latter case, the spectral weight $A_p(\omega)=z\delta(\omega)$
is  simply a delta function  at
the Fermi edge  indicating the presence of well defined
quasi-particle excitations of weight $z$ at zero temperature. At finite temperature or
finite $q$, there is the usual broadening of the quasi-particle peak ($\sim T^2$ or $v_F^2q^2$). 

The progress made to achieve  the Fourier transform of $G_p(x,\tau\to it)$
at zero temperature,\cite{Voit93,Meden92} indicates instead the absence
of quasi-particle peak in the spectral weight. In
effect,  collective modes suppresses the delta function,
which  is replaced at not too large $\theta$ by power law
singularities
\begin{eqnarray}
A_+(k_F+q,\omega)\!\! &\sim_{\omega\to u_\rho q} & \mid\!\omega 
-u_\rho q\mid^{\dm\theta  -\dm}\cr
                     &\sim_{\omega\to u_\sigma q} & 
\Theta(\omega-u_\sigma q)(\omega -u_\sigma
q)^{ \theta  -\dm}\cr
                      &\ \ \sim_{\omega\to -u_\sigma q} & 
\Theta(-\omega-u_\rho q)(-\omega
-u_\rho q)^{  \dm\theta  },
\end{eqnarray}
  for $q>0$.\cite{Voit93} At finite 
temperature,\cite{Nakamura97,Orgad01} thermal broadening will round 
 singularities and cusps.

Another physical quantity of interest is the  one-electron density of states
(per spin)  
\begin{eqnarray}
N(\omega) &= &  \sum_p\int{dq\over 2\pi}\, A_p(q,\omega)\cr
             &\propto &  \mid \omega\mid^{\theta}.
\label{density}
\end{eqnarray}
In a Luttinger liquid, the density of  states  is not constant but presents a dip close to the Fermi 
level.\cite{Suzumura80} Strickly
at the Fermi level, the density of
states is zero showing once again the absence of quasi-particles at $T=0$. At finite
temperature, $N(T)\propto
T^\theta $, the dip   partly fills at the Fermi level due to thermal
fluctuations.
\subsubsection{Two-particle response}
The two-particle response function in Matsubara-Fourier space is defined by
\begin{equation}
\chi(q,\omega_m) =\int \!\! \int  dx d\tau \,\chi(x,\tau) \ 
e^{-iqx+i\omega_m\tau},
\end{equation}
where $\chi(x,\tau)=-\langle T_\tau  O(x,\tau)O^\dagger\rangle $ is the 
two-particle correlation function. At small
$q$ and
$
\omega$, the dynamic magnetic susceptibility (or compressibility) can 
be   calculated using
the spin (charge) operator $O= (\sigma_+ +\sigma_-)/\sqrt{2}$ ($O= (\rho_+ 
+\rho_-)/\sqrt{2}$) and
  (\ref{bosonization}), with the result after analytic continuation
\begin{eqnarray}
\chi_\nu(q,i\omega_m \to \omega +i0^+) =  -{1\over \pi  u_\nu} \sum_p  {p u_\nu q\over \omega
- pu_\nu q} + i\pi \sum_p q\delta(\omega -
pu_\nu q)
\label{uniformki}
\end{eqnarray}
for both branches at zero temperature. The simple pole structure of the real part of this 
expression is
analogous to the one found for acoustic
phonons. Correspondingly, the  absence of damping shown by the imaginary
part emphasizes once more  that
spin and charge acoustic excitations are   eigenstates of the
system.   In the static ($\omega=0$)  and
uniform ($q\to0$) limits,
$\chi_\nu
\to 2(\pi u_\nu)^{-1}$. A non-zero susceptibility at zero temperature 
  then occurs despite  the absence
of  density of states at the Fermi level.  The proportionality between
the uniform response and the density of states that holds for a Fermi liquid is
 meaningless for  a Luttinger liquid due to the absence of quasi-particles.
The finite uniform response  rather probes the density 
of states of
acoustic  boson modes in
the spin or charge channel.    At non zero temperature for the TL model,
$\chi_\nu(T)$ is  only very  weakly temperature dependent on the scale of
the Debye energy
$\omega_{\nu}=u_\nu\alpha_{0}^{-1} $ of acoustic modes, which is of the
order of the Fermi energy for not too large couplings.\cite{Nelisse99,Metzner93}

Other   quantities like   staggered
density-wave (close to wave vector
$2k_F$)  responses are also of practical importance in the analysis of X-ray   and NMR
experiments.\cite{Pouget96,Pouget96b,Bourbon93,Wzietek93}  In the  following we will  focus on 
spin-spin correlation function for
  $q\sim 2k_F$; the latter can  be evaluated using the spin-density operator $ 
\vec{O}=\psi_-\vec{\sigma}\psi_+$ in the
definition of the two-particle correlation function given above. At equal-time 
for example, one gets
    the power law decay
\begin{eqnarray}
    \chi(x) &= &  \langle \vec{O}(x)\cdot \vec{O}(0)\rangle \cr
    &  \sim & \ {\cos(2k_F   x)\over
x^{1+K_\rho}},
\label{decayAF}
\end{eqnarray}
    which is    governed by the LL  parameter $K_\rho$.  The temperature dependence of the
antiferromagnetic response is given  by the
   Fourier transform of $\chi(x,\tau)$ evaluated at $2k_F$ and  in the static limit
\begin{equation}
\chi(2k_F,T) \sim T^{-\gamma}.
\label{responseAF}
\end{equation}
    The power law exponent $\gamma=1-K_\rho >0 $ is non universal and increases
with the strength of
interactions up to its highest value $\gamma=1$ 
     corresponding  to the Heisenberg universality class. A similar
expression is found for the
$2k_F$ charge-density-wave response in the TL model. For atttractive 
couplings, a power law singularity
is to be found in the
superconducting channel, where
$K_\rho$ in (\ref{decayAF}-\ref{responseAF}) is simply replaced by
$1/K_\rho$.

The imaginary part ${\rm
Im}\chi(q+2k_F,
\omega)$ of the dynamic response is another related quantity that 
plays  an important part  in
experimental situations giving an experimental access
to the Luttinger liquid parameter $ K_\rho$.\cite{Pouget96b,Wzietek93}  In the
   spin-density-wave channel of a Luttinger liquid at non zero temperature, one has the  power law enhanced
form 
      \begin{equation}
{\rm Im}\chi(q+2k_F,\omega) \sim (\pi u_\sigma)^{-1} {\omega\over T}
\left(T\over
E_F\right)^{-\gamma} 
\label{ki2kf}
\end{equation}
for small (real) frequency and $q$ close to 0.  We  shall revert 
to this below in the  context of NMR.

\subsection{The one-dimensional electron gas model}
\label{1Deg}

    When large momentum transfer ($\sim 2k_F$)  is allowed for   scattering events,  we have  an additional
coupling parameter which is    the backscattering process  denoted    $g_1$. Moreover, when the band is
half-filled (one electron per lattice site),
$4k_F$  
 coincides with the reciprocal
lattice vector $G=2\pi/a$ ($a\sim \alpha_0$ is the lattice constant) and
Umklapp scattering becomes possible.
Another coupling    
$g_3$  is then added to the  Hamiltonian  for which two electrons are
transferred from one side of the Fermi
surface to the other. The total Hamiltonian, known as
the 1D fermion gas problem now becomes
$H= H_{TL} + H'$, where
\begin{eqnarray}
    &H'&=  {1\over L}\sum_{\lbrace k,q,\alpha \rbrace}g_1\
a^{\dagger}_{+,k_1+2k_F+q,\alpha}a^{\dagger}_{-,k_2-2k_F-q,\alpha'}
a_{+,k_2,\alpha'}a_{-,k_1,\alpha} \cr
 &+&{1\over 2L}\!\!\sum_{\lbrace p,k, q,\alpha \rbrace}g_3\
a^{\dagger}_{p,k_1+p(2k_F + q),\alpha}a^{\dagger}_{p,k_2-p(2k_F +
q)+pG,-\alpha}
a_{-p,k_2,-\alpha}a_{-p,k_1,\alpha} 
\label{g1g3}
\end{eqnarray}
corresponds to the additional  terms expressed in the fermion representation.
In terms of phase variables, the part for antiparallel spins reads
\begin{eqnarray}
H' =   \int dx \,\Big\{{2g_1\over (2\pi\alpha_0)^2}
\cos(\sqrt{8}\phi_\sigma) +
    {2g_3 \over (2\pi\alpha_0)^2} \cos(\sqrt{8}\phi_\rho) \Big\},
\label{hamiltboson}
\end{eqnarray}
 whereas the parallel part of $g_1$ goes into an additional renormalization of $u_\sigma$ and $K_\nu$ in
$H_{TL}$.  From this expression, one first observes that  
$g_1$ ($g_3$)  solely depends on the spin (charge)
phase variable. Therefore the spin and  charge parts of the total Hamiltonian $H = H_\sigma + H_\rho $ 
still commute and  thus preserve   spin-charge separation. 

An exact solution of $H$ cannot be found in the general case,  except at a particular value   of the coupling
constants for each sector, corresponding to the Luther-Emery solutions \cite{LutherEmery74,Emery79}.
However, one can seek an   approximate solution  using  scaling theory.  In the framework of the
Tomanaga-Luttinger model, we have  already
emphasized  that anomalous dimensions found in single
and pair correlation functions
are hallmarks of scaling. In effect, a Luttinger liquid is a
self-similar  system when it is looked at
different length $x$ (and $u_\nu \tau $) scales. We can profit by this property for the
more general Hamiltonian by looking at
the evolution or the  flow of the couplings $g_{1,3}$  and  
Luttinger liquid parameters as a
function of successive change of    space and
time scales. In practice, the  corresponding space-time variations of the
phase variables $\phi_\nu$
are  integrated out using $g_{1,3}$ as perturbations. Then by rescaling   both the
initial length ($x\to xe^{-l}$) and time $(u_\nu\tau \to u_\nu\tau e^{-l})$  scales yields
the renormalization group flow equations
\begin{eqnarray}
{dK_\nu\over dl}= &&  - \dm K_\nu^2 g_\nu^2\cr
{dg_\nu\over dl} = && g_\nu (2-2K_\nu)
\label{Flowbose}
\end{eqnarray}
for   spin and charge parameters, where $g_\sigma\equiv g_1$ and $g_\rho
\equiv g_3$. In the
repulsive case, where $g_{i=1\ldots 4}>0$, $g_1$  is   marginally irrelevant in  the spin sector, that is to
say 
$g^*_1\to 0$ when $l\to \infty$. For a
rotationally invariant system, we have $K_\sigma^*\to 1$ and $u_\sigma^*\to
v_\sigma=v_F(1-g_4/2\pi v_F)$.       If the band filling
is incommensurate,
$g_3=0$ and we recover in this repulsive case  the physics of  a Luttinger liquid for both spin
and charge at large distance. At
half-filling, however,
$g_3$ is non zero at $l=0$ and  becomes a marginally relevant coupling  that
scales to large values as
$l$ grows; in turn, the charge stiffness $K_\rho \to 0$  and velocity
$u^*_\rho \to 0$ at large $l$. Strong
coupling in
$g_3$ and vanishing
$K_\rho$ signals the presence of a charge gap,\cite{Schulz94} which is given by
\begin{equation}
\Delta_\rho \sim E_F\left(g_{3}\over E_F\right)^{1/ [2(1-n^2K_\rho)]}, 
\label{gap}
\end{equation}
where $n=1$ at half-filling. The physics here corresponds to the one of a 1D Mott insulator.\cite{Lieb68}
The presence of a gap is also
confirmed by the fact that when
$l$ increases
$K_\rho$ decreases and the combination of
$2g_2(l)-g_1(l)$ will  invariably crosses the
so-called Luther-Emery line at
$2g_2(l_{LE})- g_1(l_{LE})=6/5$, where an exact diagonalization of the
charge Hamiltonian $H_\rho$ can be
carried out.\cite{LutherEmery74,Emery79} 

A charge gap is not limited to half-filling but may be present for other commensurabilities
too.\cite{Schulz94} At quarter-filling for example, the transfer of four particles from one side of the Fermi
surface to the other leads (instead of the above $g_3$ term) to the Umklapp coupling
\begin{equation}
H_{1/4} \simeq   {2g_{1/4}\over (2\pi\alpha_0)^2} \int dx \cos(2\sqrt{8}\phi_\rho).
\end{equation}
The scaling dimension of the operator $e^{i2\sqrt{8}\phi_\rho}$ is now $8K_\rho$, while each term of $H_{TL}$ in
(\ref{phaseH}) has a scaling dimension of 2,  so that the flow equation  becomes 
\begin{equation}
{dg_{1/4}\over dl} =  (2-8K_\rho)g_{1/4},
\end{equation}
which goes to strong coupling if $K_\rho< 1/4$, which corresponds to sizeable couplings with  longer spatial 
range.\cite{Mila93} The value of the insulating gap is given by (\ref{gap}) by taking  
$n=2$ at quarter-filling.\cite{Schulz94,Mila93} It worth noting that in the  special situation where the
quarter-filled chains are weakly dimerized, both half-filling and quarter-filling Umklapp are present with
different bare amplitudes -- a situation  met in some charge transfer salts.\cite{Barisic81,Giamarchi97}

Let us look at the consequences of a charge gap  on  correlation functions. In
the single-particle case, we see that taking $K_\rho \to 0$ at $T<
\Delta_\rho$ yields large 
$\theta_\rho$. Therefore  the
single electron coherence will become
vanishingly small at large distance. Each electron  is  confined  within
the characteristic  length scale
$\xi_\rho
\sim  v_\rho/\Delta_\rho$, which can be seen as the size of   bound
electron-hole pairs of the Mott insulator.
A large value of $\theta$ alters the spectral properties by producing a gap
in the spectral weight and in turn
the density of states. \cite{Voit98,Carlson00}

The impact is different on spin-spin correlation functions. We first note
that the uniform magnetic
susceptibility, which  uniquely depends on the spin velocity,  remains {\it
unaffected} by the charge gap, as a
consequence of the spin-charge separation (see \S~\ref{RG}). As regards
    $2k_F$ antiferromagnetic spin correlations, their amplitude increases and shows 
a slower spatial decay $\chi(x)\sim
    1/x$; this corresponds  to a stronger power law singularity in temperature
$\chi(2k_F,T)\sim T^{-1}$.

\subsection{A many-body renormalization group approach}
 Having described the basic properties of the electron gas from the
bosonic standpoint, we can now proceed to its renormalization group description from the many-body point of
view. Although the latter  works well in weak coupling, it gives a different depth of perspective in the
one-dimensional case and it proves particularly useful    in  the complex description of instabilities of
one-dimensional electronic states   when interchain coupling   is included.
\subsubsection{Renormalization group}
\label{RG}
When we try to analyze the properties of the 1D electron gas using   perturbation theory, we are faced with
infrared singularities.  These correspond to   the logarithmically  singular responses    $\chi^0\sim \ln E_F/T $
of  a free electron gas to   Cooper  {\it and }
 $ 2k_F$ electron-hole (Peierls) pair formations.   One dimension is special in that   
both share the same phase space. \cite{Bychkov66}
Their  presence with different phase relations in the electron-electron
scattering  amplitudes indicates that   
       both pairings counterbalance   one another by interference to ultimately yield a 
Luttinger liquid in leading order.

Another property of Cooper and Peierls logarithmic divergences   is the 
lack  of   particular energy scale in the  interval between
$E_F$ and $T$, a feature that  signals  scaling.
            Renormalization group ideas can  be applied to the many-body formulation in order to obtain  
the low-energy properties of the electron gas model.\cite{Solyom79,Bourbon91}   In the following, we  briefly
outline the    momentum shell Kadanoff-Wilson approach developed in
Refs.\cite{Bourbon91,Bourbon95,Bourbon02}. The partition function
$Z$ is first expressed 
as a functional integral over the fermion (Grassman) fields $\psi$
\begin{eqnarray}
Z=  \int\!\!\! \int D\psi^{\ast}D\psi \ {\rm e}^{S^*\lbrack
\psi^{\ast},\psi \rbrack}.
\label{partition}
\end{eqnarray}
     In
the Fourier-Matsubara space, the
action
$S =S_0 + S_I$  consists of a free part
\begin{eqnarray}
S_0[\psi^*,\psi] =   \sum_{p,\alpha,\kktil} 
[G^{0}_{p}(\kktil)]^{-1}\psi^{\ast}_{p,\alpha}(\kktil)\psi_{p,\alpha}(\kktil) 
\end{eqnarray}
where
\begin{equation}
G^0_p({\kktil})= \lbrack i\omega_n - \epsilon_p(k) \rbrack^{-1}
\end{equation}
is the bare electron propagator with $\kktil=(k,\omega_n)$; and an interacting part
\begin{eqnarray}
 S_I[\psi^*,\psi]=   -{T\over 2L}\!\! \!\!\sum_{\{\alpha,p,\kktil\}}\!\!
g^{\alpha_1\alpha_2;\alpha_3\alpha_4}_{p_1p_2;p_3p_4}\,
\psi^*_{p_1,\alpha_1}(\kktil_1)\psi^*_{p_2,\alpha_2}(\kktil_2)
\psi_{p_3,\alpha_3}(\kktil_3)\psi_{p_4,\alpha_4}(\kktil_4),
\label{action}
\end{eqnarray}
 in which the 
 couplings constants of the electron
gas model   are
$g^{\alpha\alpha';\alpha'\alpha}_{+-;+-}=g_1$,
$g^{\alpha\alpha';\alpha'\alpha}_{+-;-+}=g_2$,
$g^{\alpha\alpha';\alpha'\alpha}_{\pm\pm;\mp\mp}=  g_3$, and
$g^{\alpha\alpha';\alpha'\alpha}_{\pm\pm;\pm\pm}=2g_4$.\cite{Dzyaloshinskii72} The relevant parameter space of the
action for the electron gas will be denoted 
\begin{equation}
\mu_S=(G_p^0,g_1,g_2,g_3,g_4).
\end{equation}

The RG tool is used to look at the influence of high-energy states on the
electron-electron scattering near $\pm k_F$  at low
energy. We will focus on the RG results at the one-loop level,
which will be sufficient for 
our purposes. The method
consists of successive partial integrations of fermion degrees of freedom 
   ($\bar{\psi}^{(*)}$)  in the
outer energy shell (o.s)
$\pm E_0(\ell)d\ell/2$ above and below the  Fermi
points as a function of $\ell$.\cite{Bourbon91} Here  $E_0(\ell)=E_0
e^{-\ell}$ with
$\ell >0$, is   the scaled bandwidth  cutoff  $E_0  (\equiv
2E_F)$ imposed to the spectrum (\ref{Luttinger}). We can write
\begin{eqnarray}
Z  &  \sim & \int\!\!\!\int_< D\psi^*D\psi\, e^{S[\psi^*,\psi]_\ell}
\int\!\!\!\int_{\rm o.s} D\bar{\psi}^*D\bar{\psi}\,
e^{S_0[\bar{\psi}^*,\bar{\psi}]}\
e^{S_{I,2} \, +\, \ldots} \cr
  &\propto& \int\!\!\!\int_< D\psi^*D\psi\, e^{S[\psi^*,\psi]_\ell\  + \
{1\over 2} \langle
S^{2}_{I,2}
\rangle_{\rm o.s}\  + \ \ldots}\ \ \ ,
\label{trace}
\end{eqnarray}
where $S_{I,2}$ is given by the interaction term with two
$\bar{\psi}^{(*)}$ in the outer
momentum shell in the Cooper and Peierls channels ($2k_F$ electron-hole and zero momentum Cooper pairs), while
the  other two remain fixed in the inner
($<$) shell.
   
At the  one-loop level, the
averages $\langle (S_{I,2})^2\rangle_{\rm o.s}$ in the outer momentum shell
are calculated with
respect to
$S_0[\bar{\psi}^*,\bar{\psi}] $, which  ultimately leads to the scaling
transformation  of
$\mu_S$ as a function of $\ell$
\begin{equation}
{\cal
R}_{d\ell}[\mu_S(\ell)]= \mu_S(\ell +d\ell).
\label{flowtransf}
\end{equation}
The outer momentum shell contributions to the Peierls and
Cooper channels have different signs  
and lead
to the aforementioned interference  in the renormalization
flow, which is governed by the following
set of equations
\begin{eqnarray}
&& {d \tilde{g}_1\over d\ell}   = 
-\tilde{g}^{2}_1 \,+ \ldots\cr
 &&{ d  \over d\ell}(2\tilde{g}_2 -\tilde{g}_1)   =    \ \tilde{g}^{2}_3 \,+ \ldots\cr
&&{d\tilde{g}_3 \over d\ell}  =     \ \tilde{g}_3
(2\tilde{g}_2 -\tilde{g}_1)\, + \ldots,
\label{flowcouplings}
\end{eqnarray}
 where the influence of $g_4$ has been included through  the normalization
$\tilde{g}_1=g_1/\pi v_\sigma$,
$2\tilde{g}_2-\tilde{g}_1=(2{g}_2-{g}_1)/\pi v_\rho$, and
$\tilde{g}_3=g_3/\pi v_\rho$. These scaling equations,\cite{Dzyaloshinskii72}
 are
consistent with those obtained in (\ref{Flowbose}) from the bosonization
technique by expanding the   stiffness
constants
$K_{\nu=\sigma,\rho}$ to leading
order in the coupling constants. Therefore from the many-body 
 standpoint, the
interference between the Cooper and Peierls channels appears as an indispensable    
building block of  Luttinger   
 and   Luther-Emery liquids.

\subsubsection{Magnetic susceptibility}
The  description sketched above can also be used  for the calculation of
uniform responses   at small $
q$ and $\omega $  when the  
couplings
$g_1$ and
$g_3$ are present.\cite{Bourbon93,Dzyaloshinskii72,Nelisse99}  We will be mainly concerned here with the  spin
susceptibility (a similar approach also applies for  compressibility). We will see  that the flow of
$g_1(\ell) $ is responsible for  a temperature dependence of susceptibility.
 
\begin{figure}
\centerline{\includegraphics[width=7cm]{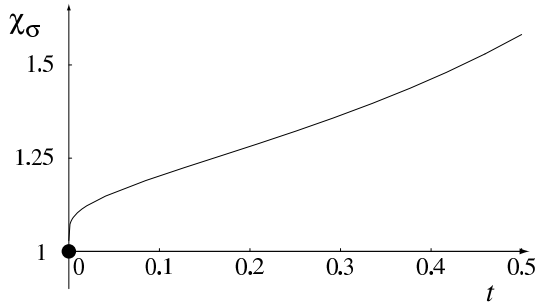}}
\caption{The temperature
variation  of the magnetic susceptibility expressed in $\pi v_\sigma/2$
units as a function of the reduced
temperature $t=T/\omega_\sigma$ for the electron gas model
($\tilde{g}_1\sim 1)$. }\label{KivsT}\end{figure}

      The first thing that needs to be said is  that thermally excited  spin
excitations  involved in  the uniform   magnetic response     do not 
contribute to the logarithmic singularities of the Cooper and Peierls
channels.
In effect, these last singularities refer  to electron and hole states located
outside the thermal
width
$\sim 2T$ around the Fermi points, while it is the other way around for the uniform spin
response of the Landau channel. The fact that the Landau
channel does not interfere directly with the other two
constitutes an advantage in the calculation. One can indeed use the
renormalization group
method to
first integrate quantum degrees in the interfering Cooper and Peierls channels,
namely  down to   $\ell_T=\ln E_F/T$, after which the resulting low-energy
action can be used to
calculate uniform  spin susceptibility.\cite{Bourbon93,Nelisse99}
If we try to outline this way of doing, we first note that the interacting
part of the action at
$\ell$ can be written 

\begin{equation}
S_{I}=\sum_{p,\qtil}(2g_{2}-g_{1})(\ell)\rho _{p}(\qtil)\rho
_{-p}(-\qtil)-g_{1}(\ell)\sum_{p,\qtil}\vec{S}_{p}(\qtil)\cdot \vec{S}_{-p}(-\qtil) +
S_I[g_4], 
\label{LandauH}
\end{equation}
in which we have defined the composite fields 
$ \rho_{\pm }=\dm(L)^{-\dm} \sum_{\kktil^*,\alpha}\psi^*_{\pm
,\alpha } \psi_{\pm,\alpha } $ for charge and $\vec{S}_{\pm } =\dm(L)^{-\dm}
\sum_{\kktil^*}\psi_{\pm ,\alpha }^{* }{\vec{\sigma}}^{\alpha \beta }\psi_{\pm,\beta } $ for spin. Here the
sum on $k$   satisfies $\mid\!\epsilon _{p}(k)\mid \leq E_{0}(\ell)/2 $; we have omit the $g_3$ term
 since it gives no direct contribution to the Landau channel. The above expression for the action
being quadratic in spin and charge fields, it  can be linearized using an
Hubbard-Stratonovich transformation which allows us to express the partition function as a functional
integral over auxilliary charge  
$\phi$ and  spin $\vec{M}$ fields 
   \begin{eqnarray}
Z    =   Z(g_4)\int\!\! \int {\cal D}{\phi}{\cal D}\vec{M}&\exp &\Big\{ - \sum_{%
\widetilde{q},p,p\prime }[\phi _{p}(\widetilde{q})A_{p,p\prime }(\widetilde{q%
})\phi _{p\prime }(-\widetilde{q})\cr
& & \hskip .5 truecm  + \  \vec{M}_{p}(\widetilde{q})B_{p,p\prime }(%
\widetilde{q})\vec{M}_{p\prime }(-\widetilde{q})] \Big\},
\label{ModeMode}
\end{eqnarray}
where $Z(g_4)$ is the partition function of the system with $g_4$
interaction only, which can be treated in RPA in the spin and charge sectors.\cite{Metzner93} The effective
low-energy free energy density   is thus essentially quadratic in both $\phi$ and $\vec{M}$
-- mode-mode coupling terms vanish   for a linear spectrum -- and can be seen as an approximate harmonic
representation of the electron gas model at low energy. From the expressions of matrix elements
$A_{p,p}(\widetilde{q})=\dm (2\tilde{g}_{2}-\tilde{g}_{1})(\ell)$, $A_{\pm,\mp}
=1$ and
$B_{\pm,\pm}(\widetilde{q})= -\dm\tilde{g}_{1}(\ell)$, $B_{\pm,\mp}=1$, the
uniform magnetic response, when expressed as statistical averages over auxiliary fields,  is given by
\begin{eqnarray}
\chi_\sigma(\widetilde{q})
&=  & \frac 1{g_1(\ell)}[\langle {\frac{1}{6}} \sum_{p,p'}\vec{M}_{p}(\widetilde{q}%
)\cdot\vec{M}_{p^{\prime }}(-\widetilde{q})\rangle -1]\cr
&=  &   -{2\over \pi} {1\over \bar{u}_\sigma(\ell) }
{u^2_\sigma(\ell)q^2\over [\omega - u_\sigma(\ell)q][\omega +
u_\sigma(\ell)q]} \, \cr
&\  & \hskip 1truecm + \  i\, {1\over  \bar{u}_\sigma(\ell) } \sum_p
u_\sigma(\ell)q\, \delta(\omega
-pu_\sigma(\ell) q),
\label{kiunif}
\end{eqnarray}
where $\bar{u}_\sigma(\ell)= v_\sigma (1 + \dm\tilde{g}_1(\ell))$. The
spectrum of low energy acoustic spin
excitations  now becomes
\begin{eqnarray}
\omega_\sigma &=  & {u}_\sigma(\ell)\mid q\mid \cr
    &=  & v_\sigma\big(1-\tilde{g}^2_1(\ell)/4\big)^{\dm}\mid q\mid,
\label{spinKi}
\end{eqnarray}
which, owing to the presence of $g_1$, shows $\ell$ dependent corrections
with respect to the
Tomanaga-Luttinger limit (Eqn.(\ref{veloTLspin})) (a similar expression
holds for the charge spectrum following
the substitution
$v_\sigma \to v_\rho$ and $-\tilde{g}_1(\ell) \to
(2\tilde{g}_2-\tilde{g}_1)(\ell)$). The temperature dependence of
the static and uniform  spin susceptibility is obtained by putting
$\qtil=0$ and $\ell\to\ln \omega_\sigma/T$
in (\ref{spinKi}) with the result
\begin{equation}
\chi_\sigma(T) = {2\over \pi v_\sigma} {1\over  1-\dm \tilde{g}_1(T)},
\label{chiT}
\end{equation}
where
\begin{equation}
\tilde{g}_{1}(T) =\frac{\tilde{g}_{1}}{1+\tilde{g}_{1}\ln
\omega_\sigma /T}  \label{g*}
\end{equation}
is the solution of the first equation of (\ref{flowcouplings}). For repulsive couplings, the
reduction of $g_1(T)$ imparts a temperature dependence to the susceptibility
which is shown in Fig.~\ref{KivsT}.
As one can see, the logarithmic corrections make the $\chi$
approaching its $T=0$ value with an
infinite slope.\cite{Dzyaloshinskii72,Bourbon93,Nelisse99,Eggert94} This singularity in the derivative occurs only
at very low temperature and can be hard to detect in practice. Finite magnetic field or interchain
hopping between
stacks   tends to suppress  the singularity.

\subsubsection{Nuclear relaxation rate}
The  nuclear spin-lattice
relaxation rate  measured in NMR experiments is another quantity of practical importance if one tries to gain
information about  spin correlations.  It is given by the Moriya expression
\begin{equation}
T_1^{-1} = \, \mid\! A\mid^2 T \int {\chi^{\prime\prime}(\vec{q},
\omega)\over \omega}\, d^Dq,
\end{equation}
which is taken in the zero Larmor frequency limit ($\omega\to 0$) and where $A$ is proportional to the
hyperfine matrix element. Thus the relaxation of nuclear spins   
 gives in principle relevant information about the static, dynamics and dimensionality
$D$ of electronic spin correlations. This gives in turn a relatively easy access to the parameters $K_\rho$,
$\bar{u}_\sigma(T)$ and $ {u}_\sigma(T)$ of the electron gas.\cite{Bourbon89}
According to Eqns.~(\ref{kiunif}) and (\ref{ki2kf}), the enhancements of
$\chi^{\prime\prime}$ occur  at
$q\sim 0 $ and in the interval $q\sim 2k_F \pm T/v_F$ close to $2k_F$.
The integration is then readily
done to give
\begin{equation}
T_1^{-1} \simeq C_0(T)T\chi_\sigma^2(T) + C_1(T)T^{K_\rho}.
\label{Relaxation}
\end{equation}
Owing to the presence of $g_1$, we have
$C_0(T)=C_0\big(u_\sigma(T)/\bar{u}_\sigma(T)\big)^{\dm}$ and
$C_1(T)=C_1(1+ \tilde{g}_1\ln\omega_\sigma/T)^{\dm}$. As a function of
temperature, two different behaviors can
  be singled out. At   high temperature, where uniform spin correlations
dominate and the $2k_F$ ones are
small, the relaxation rate is then governed by the $C_0(T)T \chi_\sigma^2(T)$ term. In the low
temperature domain, however, $2k_F$ spin correlations
 are singularly enhanced while  uniform correlations remain  finite  so that $T_1^{-1} \sim
C_1(T)T^{K_\rho}$.
The temperature dependence of
$T_1^{-1}$ over the whole temperature range   thus contrasts
with that of a Fermi liquid where $(T_1T)^{-1}\sim {\rm cst.,}$ as found for
the Korringa law in ordinary metals.

\begin{figure}
\centerline{\includegraphics[width=7cm]{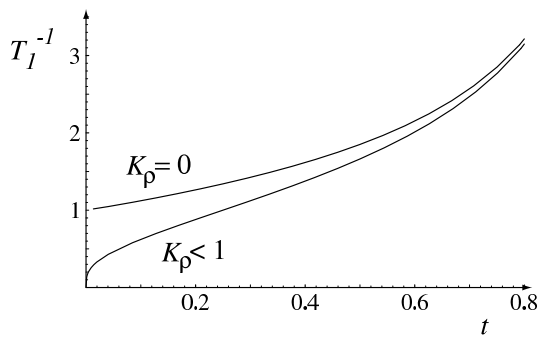}}
\caption{Nuclear spin-lattice
relaxation rate (expressed in arbitrary units) as a function of the reduced temperature 
$t=T/\omega_\sigma$, when the system scales to
the Luttinger   liquid  ($1>K_\rho >0$) and strong coupling ($K_\rho=0$)
sectors.}\label{T1theo}\end{figure}

It is interesting to consider the case where  a gap is present in the charge part and for which $K_\rho=0$. We
thus have
\begin{equation}
T_1^{-1} \simeq C_0(T)T\chi^2(T) + C_1(T).
\end{equation}
The relaxation will tend to show a finite intercept as $T\to0$ (here
logarithmic corrections in
$C_1(T)$ will give rise to an upturn in the low temperature limit).
The temperature profile is summarized
in Figure~\ref{T1theo}.

\section{Instabilities of 1D quantum liquids: interchain coupling}
\label{tperp}

Electronic materials like  organic conductors can  be only considered
as close realizations of 1D
interacting fermion systems. In the solid state,  molecular stacks 
are not completely independent,
that is to say   {\it interchain coupling}, though small,  must be
taken into account in their description.
Two different kinds of interchain coupling are generally   considered.
First, we have potential coupling like Coulomb interaction
which introduces  scattering of particles on different stacks. In
certain conditions, potential coupling may
give rise to long-range (density-wave) order at finite temperature.
Second, there is the  kinetic
coupling,  commonly denoted
$t_\perp$,   which allows an electron to hop from one stack to another.
In the following, we shall confine
ourselves to the latter coupling, which is  the most studied and by far
the most complex. In effect, it is
from
$t_\perp$ that most    instabilities of  1D quantum liquids  discussed
thus far can  occur. These
are also  called {\it crossovers} which form the   links between
the  physics  in one and higher
dimensions as either the restoration of a Fermi
liquid component or  the onset of
long-range order. In the following, we will review the physics of both
types of crossover induced by $t_\perp$   using
the renormalization group method that was described earlier.

\subsection{One-particle dimensionality crossover and beyond}
\subsubsection{ The route towards the restoration of  a Fermi liquid}
\label{Fliquid}
   The overlap of molecular orbitals allowing electrons to hop from one
chain to the next modifies the electron
spectrum. In the tight-binding picture of a linear array of 
$N_\perp$ chains we have
\begin{equation}
E_p(\kvec)= \epsilon_p(k) -2t_\perp \cos k_\perp,
\label{2Dspect}
\end{equation}
where $\kvec=(k,k_\perp)$. Generalizing  the functional-integral
representation of the partition function in
the presence of a non-zero but small $t_\perp\ll E_F$, the propagator of
the free part of the action is now
$G_p^0(\kvec,\omega_n) =[i\omega_n - E_p(\kvec)]^{-1}$. The non-interacting
situation can trivially serve to illustrate  how a dimensionality
crossover of the quantum coherence in
the single-particle
motion is achieved as a function of temperature. We first note that since
$t_\perp$ enters on the same footing as
$\omega_n$ (or
$T$) and
$\epsilon_p$ (or
$k$) in the bare propagator, it is a relevant  perturbation;   that is to
say its importance grows according to  $t_\perp'= st_\perp $ 
following the rescaling of
   energy
$\epsilon_p'=s\epsilon_p$ and temperature $T'=sT $ by a factor $s>1$. The
temperature scale at which the crossover occurs can be readily obtained by
equating $s$ with the ratio of
length scales $\xi/a$ and by setting $t_\perp'\sim E_F $ at the crossover.
This condition of isotropy yields the
crossover temperature
$T_{x^1}\sim t_\perp$, which is not a big surprise since $t_\perp$ acts as
the only characteristic energy
scale introduced in  the interval between $T$ and $E_F$.

In the presence of
interactions, however,
the flow of the enlarged parameter space $\mu_S=(G_p^0(k,\omega_n),t_\perp,
g_1,g_2,g_3)$ of the action,
under the transformation (\ref{flowtransf}) will modify this result. As
shown in great detail
elsewhere,\cite{Bourbon91,Bourbon02}   the   partial trace operation
(\ref{trace}), when carried out beyond the
one-loop level, not only alters the scattering amplitudes but also the
single-particle propagator.  At sufficiently high
energy, the
corresponding one-particle self-energy corrections   keep in first approximation their 1D character
and then modify  the purely one-dimensional part of the propagator through
the renormalization factor $
z(\ell)$. Thus the effective bare propagator at step
$\ell$  reads
\begin{equation}
G^0_p(\kvec,\omega_n,\mu_S(\ell)) = {z(\ell)\over i\omega_n -\epsilon_p(k) +
2z(\ell) t_\perp\cos k_\perp}.
\label{propagator}
\end{equation}
Detailed calculations show that $z(\ell)$ obeys a distinct flow
equation at the two-loop level
which depends on  the  couplings constants
(the generalization  of Eqn.~(\ref{flowcouplings}) at the two-loop
level).\cite{Bourbon91}
Its integration up to $\ell_T$ leads to
\begin{equation}
z(T)\sim \left({T\over E_F}\right)^{\theta},
\label{zself}
\end{equation}
where the exponent $\theta>0$ for non-zero interaction and is consistent
with the one given by the
bosonization method (cf. Eq.~(\ref{density})) in lowest order. Being the
residue at the single-particle pole
of the 1D propagator, $z(T)$ coincides with the reduction factor of the density
of states at the Fermi level (Eqn. (\ref{density})).
The  reduction of the density of states
along the chains also modifies the
amplitude of  interchain hopping. The crossover criteria mentioned
above now becomes $
(\xi/a)z(T)t_\perp\sim E_F$, which leads to the usual downward
renormalization of the one-particle
crossover temperature:\cite{Bourbon84}
\begin{equation}
T_{x^1}\sim t_\perp \left({t_\perp\over E_F}\right)^{(1-\theta)/ \theta}.
\end{equation}
According to this expression, $T_{x^1}$ decreases when the interaction $-$
which can be parametrized by
$\theta$ $-$ increases (Figure~\ref{Diagtheo}); it  is non-zero as long as
$\theta<1$  for which
$t_\perp$ remains a relevant variable. The system then undegoes a crossover
to the formation of a Fermi liquid
with  quasi-particle weight
$z(T_{x^1})$.
For strong coupling, $T_{x^1}$  vanishes  at the critical
value $\theta_c=1$ and becomes undefined for
$\theta >1$,  
$t_\perp$  being then  marginal in the former case and   irrelevant in the latter (Figure~\ref{Diagtheo}).
Consequently, in the latter case, no transverse band motion is  possible   and the
single-particle coherence is spatially confined along the
stacks. These large values of
$\theta$ cannot be attained from the above perturbative
renormalization group. They are found on
the Luther-Emery line at half-filling or at quarter-filling 
 in the presence of a
charge gap;\cite{Emery79,Voit98,Mila93} in the
gapless Tomanaga-Luttinger model for
sufficiently strong coupling constants or when the range of interaction increases.\cite{Schulz83}

The  above scaling approach to the deconfinement temperature is  obviously not
exact and  corresponds  to a random phase approximation with respect to
the transverse one-electron
motion.\cite{Bourbon84,Boies95,Arrigoni98} This can be   seen easily  by just rewriting
(\ref{propagator}) in the RPA form
\begin{equation}
G^0_p(\kvec,\omega_n,\mu_S(\ell)) = {z(\ell)G_p^0(k,\omega_n)\over 1 +
z(\ell)G_p^0(k,\omega_n) \,2t_\perp\cos
k_\perp},
\label{RPA}
\end{equation}
where $z(\ell)G_p^0(k,\omega_n)$ is the 1D propagator at the step  $\ell$
of the RG. Transverse RPA   becomes
essentially exact, however, in the limit of infinite range 
$t_\perp$.\cite{Boies95} In regard to this approximation, it   should be
stressed that the above
renormalization group treatment of deconfinement does not take into account
the dynamics
of spin-charge separation, namely the fact that the spin and charge excitations
travel at different velocities (\S~\ref{TL}). It was inferred that this
mismatch in the kinematics may
suppress $T_{x^1}$ and in turn the formation of a Fermi
liquid.\cite{Anderson91} As shown by Boies {\it
et al.}\cite{Boies95} using a functional-integral method,  the
use of the Matsubara-Fourier transform of the complete expression
(\ref{thermalG}) in the RPA allows one to
overcome this flaw of the RG. However, the calculation shows that in
the general case it still
yields  a finite crossover
temperature
\begin{equation}
T_{x^1} \approx t_\perp\left({t_\perp\over E_F}\right)^{\theta/(1-\theta)}
\left( {v_F\over
u_\rho}\right)^{\theta_\rho/(1-\theta)}\left( {v_F\over
u_\sigma}\right)^{\theta_\sigma/(1-\theta)} F[(u_\sigma/
u_\rho)^{1/2}],
\end{equation}
where $F(x)$ is a scaling function. Distinct velocities for spin and charge
do give rise to
additional corrections but the difference takes place at the quantitative
level. Electronic
deconfinement is therefore robust at this level of approximation and even
beyond.\cite{Caron88c,Schulz91,Castellani94,Arrigoni98,Biermann01b,Essler01b} 
It is worth stressing that the renormalization of $t_\perp$ by intrachain
interactions    has
been recently  confirmed on numerical grounds  for two-chain (ladder) 
fermion systems.
\cite{Capponi98,Caron01}

As we will see later, the above picture is  modified significantly
when the influence of $t_\perp$ on
pair correlations  is taken into account, especially when the amplitude of
interactions
increases and $T_{x^1}$ becomes small.
  As regards   the temperature interval over which  
one-particle crossover is achieved, it  is not expected to be very narrow. In
comparison with  crossovers in ordinary critical phenomena,   which
are confined to the close vicinity
of a phase transition,\cite{Nelson75}
deconfinement of single-particle
coherence in quasi-1D system is likely to be  spread  out over  a sizeable
temperature domain. This is so because the temperature interval $\sim [T,E_F]$ of the 1D quantum critical
domain, which is linked to the primary Luttinger liquid
fixed point, is   extremely large. Recent calculations  using   dynamical mean-field
theory seem to corroborate the existence of a sizeable  temperature  interval for the
crossover.\cite{Biermann01b}

\subsubsection{Instability of the Fermi liquid }
\label{laddersdw}

\begin{figure}
\centerline{\includegraphics[width=7cm]{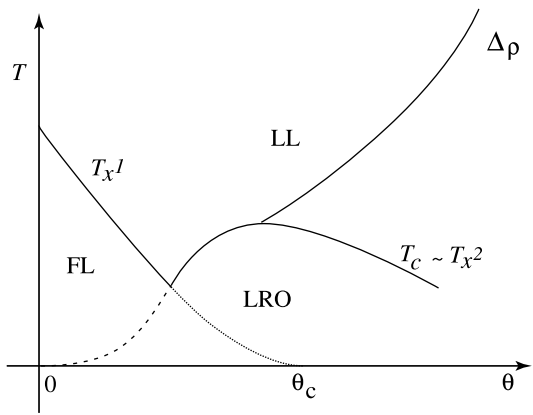}}
\caption{Characteristic temperature scales of the quasi-one-dimensional
electron gas model as a function of
interaction, parametrized by the exponent $\theta$. In the Fermi 
liquid sector, perfect nesting conditions
prevail. }\label{Diagtheo}\end{figure}

Let us  now turn  our attention to the question of whether or not a Fermi liquid
component in a quasi-one-dimensional metal
remains stable   well below
$T_{x^1}$. Here we  will 
neglect all the aforementionned transients to deconfinement  and   consider
$T_{x^1}$ as a sharp boundary  between the Luttinger and Fermi
liquids. By looking at the effective
spectrum of the above model in which
$t_\perp
\to t_\perp^*=zt_\perp$ in (\ref{2Dspect}), we observe that the 
whole spectrum obeys  
the relation
$E_-^*(\kvec)=-E_+^*(\kvec+\Qvec_0)$,
     showing electron-hole
symmetry or perfect nesting  at
$\Qvec_0= (2k_F,\pi)$.  The response of the free quasi-1D electron 
gas is still logarithmicaly singular
$\chi^0(\Qvec_0,T)\sim \ln T_{x^1}/T$ in the Peierls channel
for
$\Qvec_0$ electron-hole excitations within the   energy shell $\sim
T_{x^1}$ above and below the coherent {\it
warped} Fermi surface. This singularity is also to be found in the perturbation
theory of  the scattering amplitudes,
and   can therefore lead to an instability of the Fermi liquid. For repulsive
interactions, the most favorable instability is the one that yields
a spin-density-wave state. The  temperature at which the SDW
instability occurs can be
readily obtained by extending the renormalization group method of
\S~\ref{RG} below $T_{x^1}$ (or
$\ell >\ell_{x^1}=\ln E_F/T_{x^1}$). When perfect nesting prevails, a 
not   too bad
approximation consists of neglecting the
interference between  the Cooper and Peierls channels \cite{Prigodin79}(we shall
revert to the problem  of interference
below $T_{x^1}$ later in \S~\ref{Supra}).
Thus by retaining  the  outer
shell decomposition of
$S_{I,2}$ in the latter channel only, one can write down a ladder flow
equation
\begin{equation}
{d\tilde{J}\over d\ell} = \dm \tilde{J}^2 + ....
\end{equation}
for an effective coupling constant $\tilde{J} = \tilde{g}_2 +\tilde{g}_3 -
\tilde{V}_\perp\ldots$
that defines the net attraction between an
electron and a hole separated by $\Qvec_0$ (the origin of the exchange term  $V_\perp$
will be discussed in \S~\ref{confinement}). This equation  is
integrated at once
\begin{equation}
\tilde{J}(T) ={\tilde{J}^*\over 1-
\dm \tilde{J}^*\ln T_{x^1}/T },
\label{ladder}
\end{equation}
where
$\tilde{J}^*$ is the effective SDW coupling
at
$T_{x^1}$, resulting from the integration of 1D many-body effects at $\ell
<\ell_{x^1}$.  The above expression leads to a simple pole
singularity at the temperature scale
\begin{equation}
T_c= T_{x^1} e^{-2/J^*},
\label{BCSTc}
\end{equation}
which corresponds to a BCS type of instability of the Fermi liquid
towards   a SDW state.
   As long as the nesting conditions are fulfilled, it invariably
occurs for any non-zero interaction (dashed
line of Fig.~\ref{Diagtheo}). 

Nesting
frustration is therefore required to suppress the transition.
When nesting deviations are
sufficiently strong, we will see, however, that   the Fermi liquid is 
not stabilized
after all. Actually, when   interference between the Peierls and
   the Cooper channels is restored, the system turns out to become unstable
to superconducting pairing, a
mechanism akin to the Kohn-Luttinger mechanism for superconductivity in
isotropic
systems.\cite{Kohn65,Emery86,BealMonod86,Bourbon88,Duprat01} We shall
return to this  in \S~\ref{Supra}.

\subsection{Two-particle dimensionality crossover and pair deconfinement }
\label{confinement}
    When one examines the properties of the one-dimensional electron
gas, one observes    that the exponent
$\gamma$ of the pair response  is not simply equal to
twice   the anomalous
dimension
$\theta$ of the single-particle Green's function. Although both
exponents depend  on the Luttinger liquid
parameter  $K_\rho$, one-electron and
pair correlations are governed by distinct power law decays. Thus
the effect of an
increase in the  strength of interaction ($K_\rho $ is decreasing for
repulsive interactions)
   leads to a faster spatial decay of single particle coherence  (Eqn.~(\ref{G(x)})),
whereas the opposite is true for
triplet electron-hole pair (antiferromagnetic) correlations
(Eq.~(\ref{decayAF})).   The
    question now arises whether
$t_\perp$  can promote  interchain pair propagation besides single-particle
coherence.  Actually, this possibility exists and results
from {\it interchain pair-hopping }
processes,\cite{Bourbon86,Brazovskii85a,Firsov85} a
mechanism that is not present in the Hamiltionian at the start but
which emerges
when  interactions along
the stacks  combine  with   
$t_\perp$ in the one-dimensional region. The renormalization group approach
proved to be particularly useful in this respect making possible a
unified description of both modes of
propagation. \cite{Bourbon86,Bourbon88,Bourbon91,Bourbon95,Bourbon02}

For  repulsive interactions,
the most important pair hopping contribution is the interchain  exchange
which favors antiferromagnetic
ordering of neighboring chains.    Roughly speaking,  from each partial trace
operation in (\ref{trace}), there is a `seed'
$f(\ell)d\ell$ of interchain exchange that builds up  as a result of
combining perturbatively the 
effective hopping
($zt_\perp$) and the couplings  ($g's$) in the shell of degrees of
freedom to be integrated out. This can be seen
as a new relevant interaction for the system, which    in its turn is
magnified by
antiferromagnetic
correlations. The net   interchain exchange term generated by the
flow of  renormalization can   be
written as
\begin{equation}
S_\perp = -{1\over 4} \sum_{\langle i,j\rangle }\sum_{\qtil} V_\perp(\ell)
\,{\bf O}_i(\qtil)\cdot {\bf
O}_j(\qtil),
\end{equation}
   which favors antiferromagnetic of spins on neighboring chains $i$ and $j$.
Going to transverse Fourier space, $V_\perp$
corresponds to the exchange amplitude at the ordering wave vector
$\Qvec_0=(2k_F,\pi)$. In the one-dimensional
regime,  it is  governed at the one-loop level  by the distinct flow equation
\begin{equation}
{d\over d\ell}\tilde{V}_{\perp} = f(\ell)
+ \tilde{V}_{\perp}\gamma(\ell)
- {1 \over 2} (\tilde{V}_{\perp})^2,
\label{VRPA}
\end{equation}
where $\tilde{f}(\ell) \simeq -2[(\tilde{g}_2(\ell) + \tilde{g}_3(\ell))t_\perp/E_0]^2
e^{(2-2\theta(\ell))\ell}$. Here $\theta(\ell)$
and
$\gamma(\ell)$ are the power law exponents of the one-particle propagator
(Eqns.(\ref{density}) and
(\ref{zself}))  and antiferromagnetic response (Eqn.~(\ref{responseAF}))
respectively (these are scale dependent
due to the presence of Umklapp scattering). One observes from the
right-hand-side of the above equation  that the seed term resulting from
the perpendicular delocalization of
the electron and   hole within the pair competes with the second term due
to antiferromagnetic correlations
along the chains. The outcome of this competition will be determined 
by the sign of
$2-2\theta(\ell)-\gamma(\ell)$. Regarding  the last term, it is 
responsible for
a simple pole   singularity of
$J_\perp $ at a non-zero $\ell_c=\ln(E_F/T_c)$, signaling the onset
of long-range order at $T_c$\cite{LRO}.
The temperature at which the change from the one-dimensional regime
to  the onset of
transverse   order occurs can   be equated with a distinct
dimensionality crossover denoted by $T_{x^2} \simeq
2T_c$    for pair correlations. The latter makes sense as long as
$zt_\perp$ is still a perturbation,
that is to say for $T_{x^2} > T_{x^1}$, which defines the
region of validity of (\ref{VRPA}).

For repulsive interactions and in the presence of relevant Umklapp
scattering, one  can therefore distinguish
two different situations. The first one corresponds to the presence of  a
charge gap    well above the
transition. As we have seen earlier, it defines  a domain of $\ell$ where
$\theta(\ell)$ is large and
$\gamma(\ell)=1$, that is
$2-2\theta(\ell)-\gamma(\ell)<0$.  The physics of  this strong coupling regime
bears some resemblance to the problem of weakly coupled  Heisenberg
spin chains. However,   in the Luther-Emery
liquid case or at quarter-filling with a gap, each  electron  is not 
confined to a single site as in the
Heisenberg limit but is delocalized over
a finite          distance
$\xi_\rho
\sim v_F/ \Delta_\rho$, corresponding to  the size of bound electron-hole
pairs.  A simple analysis of
(\ref{VRPA}) shows that these pair effectively hop through an
     effective coupling $\tilde{J}_\perp\approx  (\xi_\rho
/\alpha_0)(t_\perp^{*2}/
\Delta_\rho )$. When coupled to singular correlations along the chains,
this leads to the antiferromagnetic
transition temperature
\begin{equation}
T_c \approx {t_\perp^{*2}\over \Delta_\rho}\sim T_{x^2},
\end{equation}
where $t_\perp^* = z(\Delta_\rho) t_\perp$ is the  one-particle
hopping at the energy scale of
the charge gap.

   A characteristic feature of  strong coupling
    is the increase of $T_c$  when the gap $\Delta_\rho$ decreases
(Fig.~\ref{Diagtheo}).
The above behavior of $T_c$ continues up to the point where $T_{x^2}\sim
\Delta_\rho$, namely when the
insulating behavior resulting from the charge gap merges into the critical
domain of the transition. $\theta$ and $\gamma$ take smaller values
in the normal  metallic domain
   so that   $2 -2\theta(\ell)
-\gamma(\ell)  $  will first reach zero after which it will become
positive corresponding to interchain
pair-hopping in   weak coupling.
   The  growth of the seed term then surpasses the one due to pair
vertex corrections in (\ref{VRPA}). An
approximate  expression of the transition temperature in this case is found to be
\begin{equation}
T_c \approx g^{*2}t_\perp^*,
\end{equation}
where $g^* = g_2^* + g_3^*$  and  $t_\perp^*= t_\perp z(T_c)$.  Again
this expression makes sense as
long as
$T_{x^2} > T_{x^1} $, which
on the scale of interaction should not correspond to a wide interval. Still, it
is finite and  shows  a decrease of
$T_c$ for decreasing interactions. This leads to a  maximum of $T_c$  at the
boundary between strong and weak
coupling domains (Fig.~\ref{Diagtheo}).

     As soon as $T_{x^2} < T_{x^1}$, the single particle deconfinement
occurs first at $z t_\perp\approx
E_0(\ell)
$  and interchain hopping can no longer be treated as  a perturbation.
This invalidates (\ref{VRPA}), and we
have seen earlier that  a Fermi liquid component  forms  under these
conditions. An  instability towards SDW is still
possible under good   nesting conditions for the Fermi surface. The
exchange mechanism then smoothly evolves
towards the
condensation of electron-hole pairs from a Fermi liquid. In this
regime, the residual pair-hopping amplitude
$V_\perp(\ell_{x^1})$ contributes to the effective coupling
$J^*$ in (\ref{BCSTc}). Figure 4 summarizes
the various temperature scales characterizing the
quasi-one-dimensional electron gas problem in the  presence
of Umklapp scattering.

\section{Applications}
\subsection{The Fabre and Bechgaard transfer salt  series}

\begin{figure}
\centerline{\includegraphics[width=7cm]{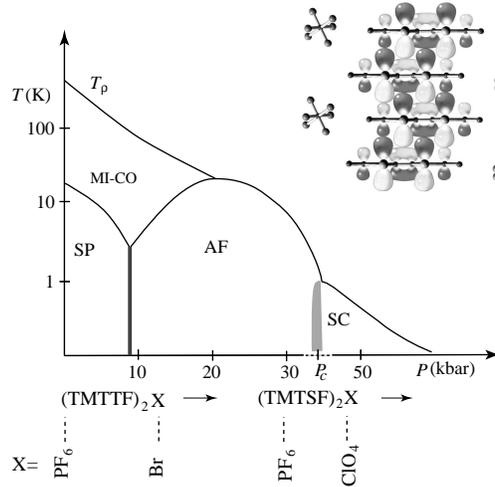}}
\caption{Temperature-Pressure phase diagram of the
Fabre ((TMTTF)$_2$X) and Bechgaard ((TMTSF)$_2$X) salts
series. Inset: a side view of the of the crystal
structure with the electronic orbitals of the
stacks.}\label{Diagexp}\end{figure}

The series of Fabre ((TMTTF)$_2$X) and Bechgaard ((TMTSF)$_2$X) transfer
salts  show striking
unity   when either hydrostatic or chemical pressure
(S/Se atom or anion $X$=PF$_6$, AsF$_6$, Br, $\ldots$, substitutions) is
applied. Electronic and structural
properties   naturally merge into the universal phase
diagram depicted in
Fig.~\ref{Diagexp}~\cite{Emery82,Caron86,Brazovskii85b,Bourbon95,Bourbon98}. 
Its
  structure reveals a characteristic sequence of ground states enabling
compounds of both series to be linked one to another
\cite{Coulon82,Brusetti82,Pouget82,Creuzet83,Wzietek93,Balicas94,Klemme95,Vescoli98,Chow98,Moser98,Chow00}.
In this way, Mott insulating  sulfur compounds like
(TMTTF)$_2$PF$_6$, AsF$_6$... were found to develop a charge-ordered (CO)
state and a lattice
distorted spin-Peierls (SP) state. The SP state is suppressed under moderate
pressure and replaced by an antiferromagnteic  (AF) N\'eel state
similar to the one found in the (TMTTF)$_2$Br salt in normal
conditions; the Mott state is in   turn suppressed under
pressure and antiferromagnetism of sulfur compounds then acquires an
itinerant character analogous to the spin-density wave (SDW) state of the
(TMTSF)$_2$X  series at low pressure.  Around some critical pressure
$P_{\rm c}$, the SDW state is then removed as the dominant ordering and
forms a common boundary with organic superconductivity which
closes the sequence of ordered states.

Within the bounds of this review, we shall not attempt a detailed discussion of the whole structure of the phase
diagram but rather place   a selected emphasis
  on the
description  of  antiferromagnetic and superconducting  orderings 
together  with their
respective normal phases.
A detailed discussion of the spin-Peierls instability and charge ordering can
be found
elsewhere.\cite{Caron87,Bourbon95,Bourbon96,Pouget96,Riera00,Chow00,Monceau01}

\begin{figure}
\centerline{\includegraphics[width=7cm]{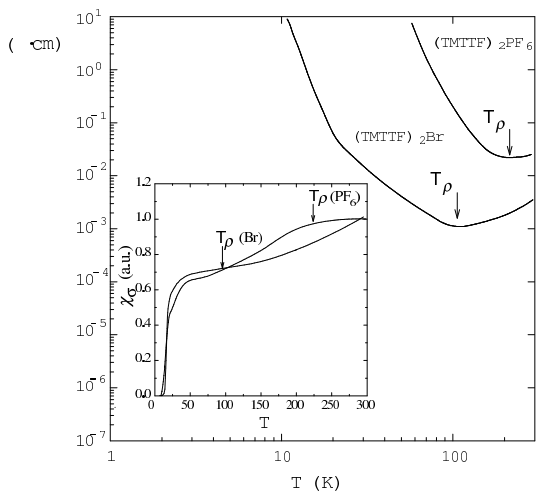}}
\caption{Temperature dependence of resistivity and static spin susceptibility
(inset) for the Fabre salts (TMTTF)$_2$Br and
(TMTTF)$_2$PF$_6$.}\label{rho-chi}\end{figure}

\paragraph{Electrical transport and susceptibility}
A convenient way to broach the description of the phase diagram is to
first examine  the normal phase
of the Fabre salts (TMTTF)$_2$X  for the inorganic monovalent anions
X= PF$_6$ and  Br.
\cite{Bourbon99}. As shown in Fig.~\ref{rho-chi},
there is a clear upturn in electrical resistivity at temperatures $T_\rho
\approx 220$~K (PF$_6$) and $T_\rho
\approx 100$~K (Br), which depicts a change from metallic to insulating
behavior.\cite{Coulon82} In both cases, $T_\rho$ is
a much higher temperature scale than the one connected to long-range 
order whose maximum
is around 20~K.   Below
$T_\rho$, charge carriers become thermally activated. In a band picture of
insulators, a
thermally activated behavior should be present for spins too.  For
the compounds shown in Fig.~\ref{rho-chi}, spin excitations  are instead
unaffected and remain gapless.
This is  shown by the regular temperature dependence of the spin
susceptibility $\chi_\sigma$ at $T_\rho$ (inset of Fig.~\ref{rho-chi}).
Resistivity data tell us that  the gap in the
charge is about $\Delta_\rho \approx 2 \ldots 2.5 T_\rho$, which exceeds
the values of $t_{\perp b}$ given by
band calculations. According to the  discussion given in
\S~\ref{confinement}, this would correspond to a
situation of strong electronic confinement along the chains. Confinement
is confirmed by the absence of a plasma
edge in the reflectivity of both compounds when the electric field  is oriented
along the transverse $b$ direction.\cite{Vescoli98,Schwartz98}

The magnetic susceptibility, which decreases with the temperature, is also
compatible with the one-dimensional
prediction of Fig.~\ref{KivsT}. Wzietek {\it et al.,}\cite{Wzietek93} 
analyzed in
detail the NMR suceptibility   data
at constant volume using the expression (\ref{chiT}). Very reasonable fits
were obtained above 50~K provided that
interactions $g_{1,2,4}   \simeq 1 $ are sizeable. In the charge
sector,  when the origin of the gap is  attributed  to
half-filling umklapp alone, one has for a compound like (TMTTF)$_2$PF$_6$ the small bare value
$\tilde{g}_3\sim  0.1$; which is obtained from reasonable band parameters   by matching the experimental
$T_\rho$  and the value   $\ell_\rho$ at which $\tilde{g}_3(\ell_\rho)\sim 1$ in (\ref{flowcouplings}).
 \cite{Caron86,Bourbon95} This is consistent  with the fact
that the stacks are weakly
dimerized.\cite{Barisic81,Emery82,Mila94} In the quarter-filling
scenario, $g_{1/4}$ would be much  larger using  small $K_\rho$ in (\ref{gap}).

\begin{figure}
\centerline{\includegraphics[width=7cm]{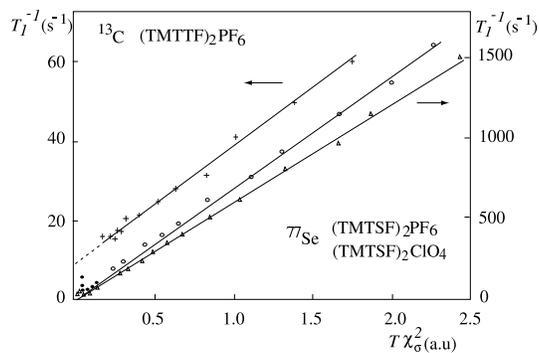}}
\caption{ Temperature dependence of the nuclear relaxation rate plotted as
$T_1^{-1} {\it
vs\  } T\chi_\sigma^2(T)$, where $\chi_\sigma(T)$ is the measured spin susceptibility.
(TMTTF)$_2$PF$_6$ (crosses, left scale), (TMTSF)$_2$PF$_6$ (open circles,
right scale) and (TMTSF)$_2$ClO$_4$
   (open triangles, right scale). 
After~Ref.\cite{Bourbon89}.}\label{T1}\end{figure}

\paragraph{Nuclear magnetic resonance}
Among other measurable quantities that are sensitive to 
one-dimensional physics,
we have the temperature dependence of
the nuclear spin-lattice relaxation rate
$T_1^{-1}$ .\cite{Wzietek93} Consider for example the insulating compounds
   (TMTTF)$_2$X. According   to the scaling pictures of  \S~\ref{1Deg} 
and \S~\ref{RG}, the charge
stiffness \hbox{$K_\rho=0$}
vanishes in the presence of  a   gap  below $T_\rho$, or
for $\ell > \ell_\rho$. The resulting behavior for the relaxation rate is then
\begin{equation}
   T_1^{-1}\sim C_1
+ C_0T\chi_\sigma^2.
\end{equation}
As shown in Fig.~\ref{T1}, this behavior is indeed found  for
(TMTTF)$_2$PF$_6$ salt when the  relaxation rate
data are combined with  those of the spin susceptibility ($T\chi_\sigma^2)$ in the normal
phase above 40~K, namely above the onset of spin-Peierls 
fluctuations.\cite{Creuzet87,Bourbon89} A similar
behavior  is  invariably found in all  insulating 
materials down to  low
temperature  where three-dimensional
magnetic or lattice long-range order is
stabilized.\cite{Wzietek93,Jerome94,Gotschy92} Long-range order also
prevents the observation of logarithmic corrections in $C_1$ which are
expected to   show up in the low
temperature limit.

If we now turn  our attention to the effect of pressure, the phase diagram
of Fig.~\ref{Diagexp} shows that
hydrostatic pressure reduces $T_\rho$. At sufficiently high pressure, the
insulating behavior merges with the
critical behavior associated with the formation of  
   a spin-density-wave 
state.\cite{Wilhelm01,Balicas94,Jerome80} The
   normal phase is then entirely metallic. This change of
behavior can also be
achieved {\it via} chemical means. We
have already seen the effect of chemical pressure within the Fabre 
salts series   when  for
example the monovalent anion Br
was put in place of PF$_6$ leading to a sizeable decrease of $T_\rho$
    (Fig.~\ref{rho-chi}). When we substitute   TMTSF    for
TMTTF, however, it
leads to  a larger shift of the pressure scale as exemplified 
by the normal
phase of  the Bechgaard salts
(TMTSF)$_2$X (X=PF$_6$, AsF$_6$, ClO$_4$, \ldots), which is metallic.
Assuming that there is a temperature domain where a one-dimensional 
picture   applies
to mobile carriers,  one must have $K_\rho >0$. Therefore
   the contribution of uniform
spin excitations to the relaxation rate becomes more important.
For a compound like  (TMTSF)$_2$PF$_6$, which
develops a  SDW state at
$T_c\approx 12$~K, deviations  to the $T\chi_\sigma^2$  law due to antiferromagetic
correlations become visible below
$T\chi_s^2\approx 1$ ($\approx 200$~K); whereas for a ambient pressure
superconductor like (TMTSF)$_2$ClO$_4$
($T_c\approx 1.2~K)$, which is on the right of the PF$_6$ salt on the
pressure scale,   deviations show up at
much lower temperature,  $T\approx 30$~K or $T\chi_s^2\approx
0.1$ in Fig.~\ref{T1}. Attempts to square these  non critical
antiferromagnetic enhancements   with the
Luttinger liquid picture, however,  show
that
$K_\rho\simeq 0.1$.\cite{Wzietek93,Bourbon84}    Figure~\ref{T1}
serves to illustrate  how the strength
of antiferromagnetic correlations decrease as one
moves from the left to the right  side of the phase diagram.

\paragraph{DC transport and optical conductivity}
The  metallic  resistivity   of the
Fabre salts ($T > T_\rho$ in
Fig.~\ref{rho-chi}) and of the Bechgaard salts at high temperature
has also been analyzed in
the one-dimensional
framework.\cite{Giamarchi97} It was shown that    Umklapp
scattering at quarter-filling is the only  mechanism of
electronic relaxation that can yield metallic resistivity above $T_\rho$
$-$ half-filling Umklapp alone would
lead at small $K_\rho$ to   an insulating
behavior at all  temperatures.\cite{Gorkov73}  Following
Giamarchi
\cite{Giamarchi97,Giamarchi92}, the  prediction at quarter-filling
is  
$\rho(T) \sim T^{16K_\rho-3}$,
    which  can reasonably
account for the constant volume metallic resistivity observed  at high
temperature.  In the Bechgaard salts  for example,  an
essentially linear  temperature dependence
$\rho(T)
\sim T ^{\nu}$  with $\nu\simeq 1$ is found
down to $100$~K, which would correspond to $K_\rho\simeq 0.25$
\cite{Schwartz98,Bourbon99}, a value not too far from NMR estimates
for (TMTSF)$_2$PF$_6$. Below,  a
stronger power law sets in  approaching a Fermi liquid behavior with
$\nu\simeq 2$, which would indicate the onset
of electronic deconfinement.  Roughly similar conclusions, as to the value
of the charge stiffness $K_\rho$ and
the onset of deconfinement in the Bechgaard salts, have been reached
from the analysis
of DC transverse resistivity measurements
in the high temperature region.\cite{Moser98,Fertey99} These are also
characterized by a marked change of
behavior taking place between 50...100~K.\cite{Moser99}

\begin{figure}
\centerline{\includegraphics[width=7cm]{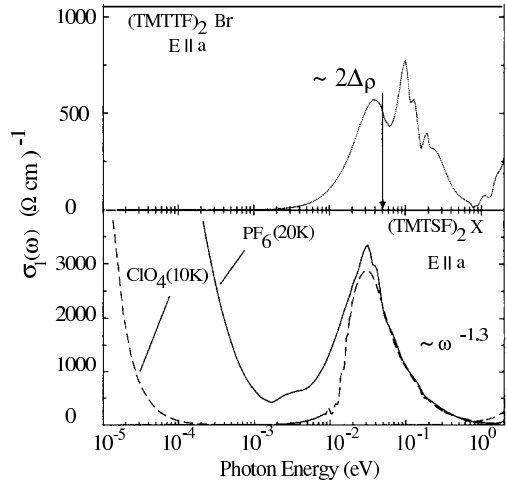}}
\caption{Optical conductivity of the Fabre salt (TMTTF)$_2$Br (top) and the
Bechgaard salts (TMTSF)$_2$PF$_6$ and (TMTSF)$_2$ClO$_4$ After
Ref.~\cite{Vescoli98}.}\label{Conduct}\end{figure}

Optical conductivity measurements on members of both
series have recently prompted a
lot  of interest  in   the extent to which a one-dimensional 
description applies to Fabre and
Bechgaard salts.\cite{Dressel96,Vescoli98,Schwartz98,Bourbon98}  As shown in
Fig.~\ref{Conduct}, sulfur compounds show the absence of a
Drude weight in the low frequency limit and the infrared  conductivity is
entirely dominated by in optical gap
of the charge sector as expected. It is noteworthy that the optical
gap is closer to 2$\Delta_\rho$ than
  $\Delta_\rho$. Following  recent work of Essler and Tsvelik,
\cite{Essler01}  this is consistent with  double
solitonic excitations in the charge sector 
 and thus a gap produced by quarter-filling Umklapp
scattering. 

 The results for
conductivity in the Bechgaard salts came as a surprise, however,
since despite the pronounced metallic
character of these systems \cite{Bechgaard80}, and the existence of a
very narrow zero frequency mode,
the charge gap still captures most of the spectral weight at high
frequencies.\cite{Vescoli98,Schwartz98,Cao96} This behavior turns out
  to mimic that of a doped Mott
insulator.\cite{Giamarchi97} According to this picture, the high
frequency tail of the conductivity above the gap   behave as
$\sigma_1(\omega) \sim
\omega^{16 K_\rho-5} $ for quarter-filling Umklapp in
1D.\cite{Giamarchi92,Controzzi01} A power law
$\omega^{-1.3}$ is  observed for  the Bechgaard salts over more than a
decade in frequency (Fig.~\ref{Conduct}), which yields the value
$K_\rho\simeq 0.23$  for the charge stiffness. This is consistent
with  the estimate made from  DC
resistivity. The purely one-dimensional prediction works well in the high
frequency range presumably because  the
effect of interchain hopping is small there ($\omega > t_{\perp b}$). However,
deviations   from what is expected
in a  1D doped Mott insulator are seen at lower frequencies  and have been
attributed to the
influence of $t_{\perp b}$.

\begin{figure}
\centerline{\includegraphics[width=5.5cm]{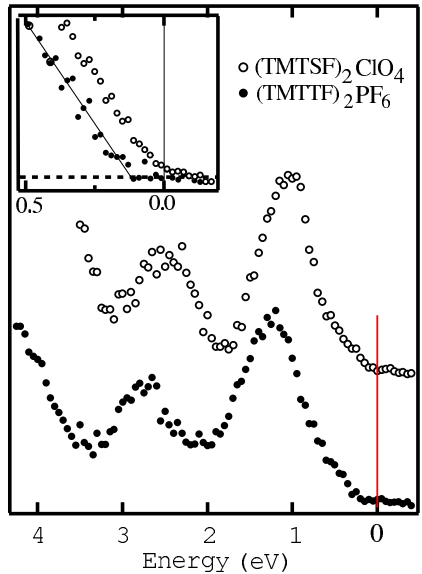}}
\caption{ ARPES spectra of (TMTTF)$_2$PF$_6$ and (TMTSF)$_2$ClO$_4$. The
inset identifies an energy shift corresponding to $2\Delta_\rho$ for
(TMTTF)$_2$PF$_6$.  After Ref.~\cite{Zwick97}.}\label{ARPES}\end{figure}

\paragraph{Photoemission results}
As mentionned earlier in \S~\ref{PropLL}, Angular Resolved Photoemission Spectroscopy (ARPES) experiments give 
in principle access to momentum and energy dependence of the one-particle spectral density $A(\vec{q},\omega)$ for
$ \vec{q} <0$\cite{Grioni00,Gweon01}. In practice, this is submitted to the experimental constraints of
energy ($\Delta\omega$) and momentum ($\Delta \vec{q}$) resolutions, and to thermal broadening. The
photoemission signal will then go  like \cite{Gweon01}
$$
I(\vec{q},\omega,T)\sim \sum_{\Delta\omega'} f(\omega')\sum_{\Delta \vec{q}'} A(\vec{q}',\omega',T),
$$
where $f$ is the Fermi-Dirac distribution. ARPES measurements of Zwick {\it et al.,}\cite{Zwick97} for the
Bechgaard salt (TMTSF)$_2$ClO$_4$ and the Fabre salt (TMTTF)$_2$PF$_6$ are shown in Fig.~\ref{ARPES}. The data are
amazing in many respects. In (TMTSF)$_2$ClO$_4$ for example, which is a 1K superconductor (Fig.~\ref{Diagexp}),
the data reveal weak spectral intensity at the Fermi level and the absence of dispersing low-energy peaks
associated  to  either   spin or  charge  degrees of freedom. The exponent $\theta $ needed to describe
the ($\sim $ linear) energy profile of $I$ down to the non dispersing  peak at $\sim 1$~eV is rather large.
Although the origin of these peculiar features is as yet not fully understood, it has been proposed that cleaving
and radiation alteration of the surface may introduce imperfections and defects that may change the properties of
the Luttinger liquid near the surface.\cite{Grioni00,Voit00} Defects can be seen as introducing  finite
segments of chains with open boundary conditions which correspond to {\it bounded Luttinger
liquids}.\cite{Voit00}  Their spectral weight is given by
$$
A(\omega)\sim \mid\omega\mid^{(2K_\rho)^{-1} -1/2},
$$  
which is $k-$independent, that is non dispersing but  with an exponent that is still governed  by the $K_\rho$ of
the bulk. Thus by taking 
$K_\rho\simeq 1/3$, which is actually not too far from the values of other experiments discused above, would lead
to power law compatible with experimental findings. 

The ARPES data for the Mott insulating  compound  (TMTTF)$_2$PF$_6$ reveal similar features apart a clear shift
$\sim 100$~mev of the onset towards negative energy. The shift seems to be close  to  $\sim
2\Delta_\rho$,  as expected for    quarter-filling Umklapp.\cite{Essler01b}

\paragraph{Dimensionality crossovers and long-range order }
Our next task will be  to give a brief description of the onset of
long-range order in  the light of  ideas  
developed earlier in \S~\ref{tperp} for an array of weakly coupled 1D electron gas.
It was pointed out that the nature of
the 1D electronic state strongly influences the way the system
undergoes a crossover  to a higher dimensional
behavior and how this may yield to long-range order. We have seen for
example  that in the presence
of a 1D Mott insulating state, the  charge
gap  gives rise to strong coupling conditions  that prevent
electronic deconfinement.
Spins can still order antiferromagnetically in
the transverse direction
{\it via} the interchain
exchange
$V_\perp$. The case of (TMTTF)$_2$Br compound is particularly
interesting in this respect since  this
mechanism can be studied in a relatively narrow
pressure interval where  the change from  strong to weak coupling
actually takes place  (Fig.~\ref{Diagexp}).

\begin{figure}
\centerline{\includegraphics[width=7cm]{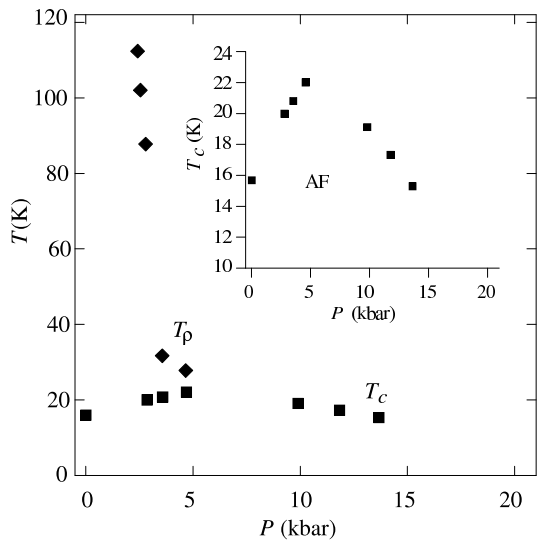}}
\caption{The pressure profile of $T_\rho$ (full diamonds) and the
antiferromagnetic critical temperature (full squares)  for
(TMTTF)$_2$Br. After Ref.~\cite{Klemme95}}\label{DiagBr}\end{figure}

The data of Fig.~\ref{DiagBr} obtained by Klemme {\it et al.,}
\cite{Klemme95} gives the detailed pressure
dependence of both $T_c$ and
$T_\rho$ for the bromine salt up to 13~kbar. When the insulating
behavior at $T_\rho$ meets  the critical domain under pressure a maximum of
$T_c$  is clearly seen. This
accords well with the
description of the transition given in \S~\ref{confinement} in terms of
weakly coupled
antiferromagnetic chains (right side of
Fig.~\ref{Diagtheo}).  Here pressure mainly contracts the lattice, modifying
upward  longitudinal
bandwidth  while decreasing the stack dimerization. These are consistent with the
decrease of correlations along the chains
whose strength has been parametrized by
$\theta$ in Fig.~\ref{Diagtheo}. Pressure  increases $t_\perp$ too -- 
roughly at the same rate as
$t_\parallel$ \cite{Jerome82}-- which   contributes to the variation of
$\theta$ under pressure.\cite{Biermann01b} Similar profiles of
$(T_c,T_\rho)$ under pressure have been confirmed in other
members of the Fabre salt  series.
\cite{Jaccard01,Wilhelm01,Adachi00,Auban-Senzier89}

Passed the maximum, the interchain exchange is    
gradually replaced by an 
instability of the Fermi surface to form 
a SDW state. We have  emphasized in
\S~\ref{confinement}, however, that this is closely related  to the issue of
  where the onset of electronic
deconfinement takes place in the  metallic state. In the general phase diagram
of Fig.~\ref{Diagexp}, this fundamental issue applies to the region where
the Fabre salt series  overlaps with the
Bechgaard salt series.  In the simple
picture  given in
\S~\ref{Fliquid}, which is portrayed  in Fig.~\ref{Diagtheo}, the scale
$T_{x^1}$ for deconfinement and restoration of a Fermi liquid component
rises  up in the small coupling region.
On an empirical basis, however, a clear indication  of such a scale
is still missing so far and the
figures proposed have led to conflicting views.      If one  agrees
for example on the small value of
$K_\rho\simeq 0.2$ given by optics and  transport 
   for the Bechgaard salts, the expected $T_{x^1} \sim 10$~K
would be  rather small. Although
this would be consistent with
earlier interpretation of NMR results, \cite{Bourbon84,Wzietek93} it 
contradicts others which favour  a Fermi liquid description: the
emergence of a
$T^2$ law  for parallel resistivity below 100~K, the gradual onset of
transverse
plasma edge in the same temperature range,\cite{Vescoli98,Jacobsen81}
the $t_\perp$ values extracted
from angular dependence of magnetoresitance a very low temperature,\cite{Danner94} the
observation of Wiedeman-Franz Law  at low
temperature,\cite{Belin97}  to mention only  a few (a more detailed
discussion can be found in
Refs.~\cite{Bourbon99,Schwartz98,Bourbon98}).  At  present, it is
not clear to what extent a synthesis of these conflicting figures  will
require a radical change of approach  in setting out  the
deconfinement problem or if it simply reflects the fact that   deconfinement
takes place over a large temperature interval.\cite{Biermann01b}

\begin{figure}
\centerline{\includegraphics[width=7cm] {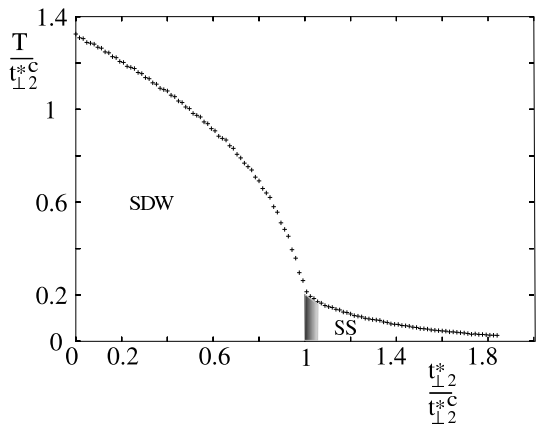}}
\caption{Variation of the critical  temperature as a function of nesting
deviations $t_{\perp2}^*$ ($\sim$ pressure). The shaded area corresponds to the crossover
region between SDW and superconductivity. After
Ref.~\cite{Duprat01}.}\label{TcSDWSS}\end{figure}

Before closing this discussion, we will briefly examine  the mechanism
commonly held responsible for the
suppression of the SDW state as the critical pressure $P_c$ is
approached from below in the phase
diagram (Fig.~\ref{Diagexp}). By looking more closely at the effect
of pressure on electronic band
structure, we realize that corrections to the spectrum such as the
longitudinal curvature of the band or
transverse hopping $t_{\perp2}$  to second nearest-neighbor chains
magnify under pressure; their
influence can no longer be neglected in the description of the  SDW
instability of the normal  state.
In effect, in the presence of an effective $t^*_{\perp2}$, the
spectrum below $T_{x^1}$ becomes 
$$
E_p(\kvec)= \epsilon_p(k) - 2t_\perp^*\cos k_\perp -
2t_{\perp2}^*\cos 2k_\perp.
$$
This leads to $E_+(\kvec + \Qvec_0) = - E_{-}(\kvec)  +   4t_{\perp2}^*\cos
2k_\perp$,  and thus to the alteration
of nesting conditions of the whole Fermi surface. These deviations  will cut off
the  infrared singularity of
the Peierls channel,  which
 becomes rhoughly $\chi^0(\Qvec_0,T)\sim
N(0)\ln(\sqrt{T^2 +\delta^2}/T_{x^1})$, where $\delta\sim
t_{\perp2}^*$. Its substitution in  the ladder
expression (\ref{ladder})  leads to a $T_c$ that rapidly  goes down
  when $t_{\perp2}^*\sim T_c$ (see Fig.~\ref{TcSDWSS} for the results 
of a detailed calculation), in
qualitative agreement
   with experimental findings (Fig.~\ref{Diagexp}).

The strongest support for the relevance of nesting
frustration in this part of the phase diagram is
provided by the analysis of the  cascade of SDW phases at $P>P_c$, 
which are observed
when a magnetic field oriented along
the less conducting direction is cranked up beyond some threshold.\cite{Chaikin96,Jerome94} In 
effect, the magnetic field confines
the electronic motion in the transverse direction and thus restores 
the infrared singularity of the Peierls
channel at some discrete (quantized) values of the nesting vector, each of
which characterizing a SDW phase of the
cascade.\cite{FISDW} Besides the indisputable success of a weak coupling
(ladder) description of
field-induced SDW $-$ which constitutes a whole chapter of the physics of
these materials \cite{Ishiguro90}$-$
  some features of the normal phase under field such as the magnetoresistance
and NMR  refuse to bow   to a
simple Fermi liquid description.
\cite{Behnia95}

\paragraph{On the nature of superconductivity}
\label{Supra}
Let us now turn our attention to the superconductivity that is found  near $P_c$ for 
both series, namely
at the right hand end of the phase diagram
in Fig.~\ref{Diagexp}. The symmetry $-$ singlet or triplet $-$ of the 
superconducting order parameter in these charge-transfer  salts is
an  open question that is currently much debated\footnote{See the review of P. M.
Chaikin in this volume.}.\cite{Lee97,Lee02,Lebed99}  Here we will  tackle
this problem   from a  theoretical standpoint that is in line with 
what has been previously discussed.
    We  shall  give  a cursory glance at  recent progress made on  the  origin of organic 
superconductivity, namely as to whether it
could be driven by electronic correlations.   In this matter, it is 
noteworthy that superconductivity
shares a common boundary and even overlaps with the SDW
state.\cite{Jerome80,Wilhelm01,Adachi00,Brusetti82} This close 
proximity between the two ground states,
which is an universal feature of both series of compounds, is 
peculiar in  that  it is the electrons of a
single band that  partake in both types of long-range order. 
Moreover, SDW correlations are well known to
permeate deeply the normal phase above the superconducting 
phase.\cite{Wzietek93} All this goes to show
that pairings between electrons and holes responsible for 
antiferromagnetism and superconductivity are not
entirely exclusive
  and that both  phenomena may have a common $-$ electronic $-$ origin.
\cite{Duprat01,Emery86,Caron86,BealMonod86,Bourbon88}

We have  become  familiar with the mixing of Cooper and Peierls
pairings in the context of a Luttinger liquid
  (see \S~\ref{RG}). The interference between the two is maximum in
strictly one dimension where  the Fermi surface reduces to two 
points. Below the scale $T_{x^1}$,   interference was
neglected  in the ladder description of the SDW instability
(see \S~\ref{laddersdw}). However,  although interference is weakened by 
the presence of a coherent
wrapping of the Fermi surface, it still exerts
an influence below
$T_{x^1}$ by  becoming non uniform, that is $\kvec$-dependent along 
the Fermi surface. As shown recently by
Duprat {\it et al.,}\cite{Duprat01}
  non uniform interference can be taken into account using the 
renormalization group method of \S~\ref{RG}.  This technique 
allows us to write down a two-variable flow equation for the SDW 
coupling constant
\begin{eqnarray}
{d\tilde{J}(k_\perp,k_\perp')\over d\ell} = & -&{1\over  N_\perp}
\sum_{k_\perp^{''}}\tilde{J}(k_\perp,k_\perp^{''})I_{C}(k_\perp^{''},k_\perp',\ell)
\tilde{J}(k_\perp^{''},k_\perp^{'}) \ \cr &+ &\ \ {1\over N_\perp}\sum_{k_\perp^{''}}
\tilde{J}(k_\perp,k_\perp^{''})I_{P}(k_\perp^{''},k_\perp',t_{\perp2}^*,\ell),
\tilde{J}(k_\perp^{''},k_\perp^{'}) 
\end{eqnarray}
where $I_{C,P}$ are the  Cooper  loop and the Peierls one in the 
presence of nesting deviations. The
pair of variables ($k_\perp,k_\perp')$ refers to  transverse momenta 
of ingoing and outgoing
particles that participate in electron-hole and electron-electron 
pairings close to the Fermi surface.

  At
small
$t_{\perp2}^*$, there is a simple pole singularity in
$J$ at
$k_\perp'=k_\perp-\pi$, which  signals the expected SDW instability 
at  $\ell_c=\ln
T_c/T_{x^1}$ and wavevector $\Qvec_0$ (Fig.~\ref{TcSDWSS}). As far as 
$T_c$ is concerned, this result is
qualitatively similar to the
   ladder approximation of  \S~\ref{laddersdw}.  When nesting deviations increase, 
however, $T_c$ decreases rapidly and shows
an inflection point at a critical $t_{\perp2 }^{*c}$  instead of 
reaching zero as for the single channel
approximation (Fig.~\ref{TcSDWSS}). The singular structure of $J$ in $k_\perp,k_\perp'$ 
space then qualitatively changes,
   becoming modulated by a product of simple harmonics $\cos 
k_\perp\cos k_\perp'$. This indicates that
singular pairing is now present in  the singlet Cooper channel. The 
attraction takes place between
electrons on neighboring stacks as a result of their coupling to spin 
fluctuations. In the framework of the
present model,  the singlet superconducting gap
$\Delta(k_\perp)=\Delta_0 \cos k_\perp $  presents nodes at 
$k_\perp=\pm \pi/2$. When typical
figures for deviations to perfect nesting are used, that is 
$t_{\perp2,c}^*\sim 10$~K $\sim
T_c(t_{\perp2}^*\approx 0)$, the one-loop calculations is able to 
reproduce  an important feature  of the phase diagram 
which is the profiles of SDW and superconductivity in both series of 
compounds near  $P_c$ (Fig.~\ref{Diagexp}).

\subsection{The special case of TTF[Ni(dmit)$_2$]$_2$}

Among the very few   
quasi-one-dimensional organic materials that do not  show  long-range
ordering, the two-chain compound TTF[Ni(dmit)$_2$]$_2$ is
interesting. \cite{Brossard86,Bourbon88b} The TTF stacks remain  metallic
down to
the lowest  temperature reached for this system. Although the reason for
this lack of long-range order is not
well understood, band calculations revealed a very pronounced quasi-1D
anisotropy of the electronic
structure,\cite{Kobayashi87} actually stronger than the one of the Fabre
and the Bechgaard salts. The band filling
is not known precisely but it is incommensurate with the underlying
lattice. We have here favorable conditions for
the emergence of Luttinger liquid physics.

\begin{figure}
\centerline{\includegraphics[width=7cm] {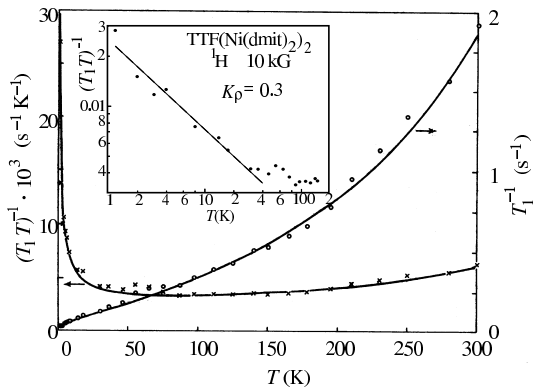}}
\caption{The temperature dependence of the nuclear spin-lattice
relaxation rate of TTF[Ni(dmit)$_2$]. The continuous line is a Luttinger
liquid fit.
The power law enhancement at low temperature is shown in the inset. After
Wzietek {\it et al.}~\cite{Bourbon88b}.}\label{T1dmit}\end{figure}

  In this respect, the results of Wzietek {\it et al.,}~\cite{Bourbon88b}
given in Figure~\ref{T1dmit} for the
temperature variation of the proton ($^1$H) NMR $T_1^{-1}$ of the TTF
chains, are  particularly revealing.
When the data of Fig.~\ref{T1dmit} are compared with the characteristic LL
shape of Figure~\ref{T1theo} at $K_\rho >0$, the connection
with the one-dimenional theory is striking. This is confirmed at the
quantitative level  by the fit (continuous
line in Fig.~\ref{T1dmit}) of data using an expression  of the form (\ref{Relaxation}), where
the interaction parameter  $K_\rho\simeq
0.3$  have been used. The plot of
$(T_1T)^{-1}$ {\it vs} $T$  (left scale and inset) allows one  to isolate
the enhancement at low temperature due
to the 1D antiferromagnetic response.

\subsection{An incursion in inorganics: the purple bronze
Li$_{0.9}$Mo$_6$O$_{17}$}
The objective  of finding   Luttinger liquid behavior in   crystals does
not   focus uniquely on organic
conductors but constitutes an important line of research in other
materials too. This is the case of
molybendum bronzes which form a class of low-dimensional  inorganic systems
well known for their strongly
anisotropic electronic structure and the  rich phenomenology associated with
the formation of
charge-density-wave order.\cite{Schlenker96}
    Here we will   briefly consider     the Li$_{0.9}$Mo$_6$O$_{17}$
compound, which
   stands out as a special case of the  so-called `purple bronze' series.
This compound consists of molecular
MoO$_6$ octahedra and MoO$_4$ tetrahedra arranged in a 3D network   for
which strong Mo-O-Mo interactions form
zig-zag chains along a preferential direction. This strong one-dimensional character is confirmed by band calculations 
\cite{Whangbo88} and by experimental Fermi surface mapping,\cite{Denlinger99} which both yield a 
rather flat Fermi surface arising from the two degenerate bands that cross 
the Fermi level.
   The metallic temperature domain is rather wide, extending down to
$T_c
\approx 24$~K where  a phase transition   occurs. Although the origin of
the latter is as yet not well understood,
it is not a CDW state and  the normal phase  does not show any sign of CDW
precursors. Therefore electronic
interactions dominate and this renders  this material particularly
appealing for ARPES studies.\cite{Denlinger99}

\begin{figure}
\centerline{\includegraphics[width=7cm] {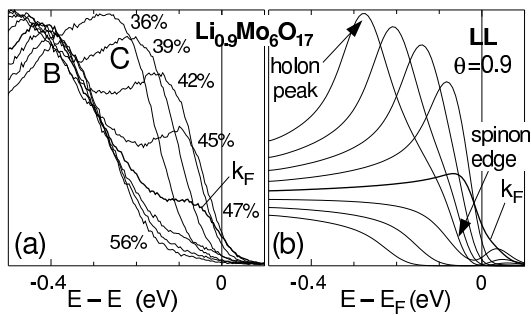}}
\caption{(a) ARPES data at
250~K for Li$_{0.9}$Mo$_6$O$_{17}$ at  different \%   of the Brillouin zone
in the $\Gamma-$Y direction  ; (b)
   Luttinger liquid prediction at $\theta= 0.9$,  and $u_\sigma=u_\rho/5$,
including
experimental momentum and energy resolutions, and thermal (Fermi-Dirac)
broadening. After
Ref.~\cite{Gweon01b}.}\label{Lithium}\end{figure}

    Figure~\ref{Lithium} shows high resolution ARPES data obtained  by Gweon
{\it et
al.,} \cite{Gweon01b} on Li$_{0.9}$Mo$_6$O$_{17}$ at $T=250$~K. As one
moves along the \hbox{$\Gamma\!-\!Y$}
direction in the Brillouin zone, the lineshape of the C band   shifts
towards the Fermi edge for increasing  ${\bf
k} < {\bf k}_F$ with a peak that decreases sharply in intensity and
broadens significantly before reaching
${\bf k}_F$ to finally pull back for ${\bf k} > {\bf k}_F$ and merges in  the
tail of a  weakly dispersing band (B)
at high energy.
   The absence of true crossing of the band, its  lack  of sharpening  as
${\bf k}
\to {\bf k}_F$  and the low spectral
weight left at {\bf k}$_F$  contrast with what is found in the pototypical
Fermi liquid like  compound  TiTe$_2$
\cite{Gweon01,Straub97,Claessen92}. The analysis of the dispersing lineshape C in the framework of the LL
theory proved to be much more satisfactory.\cite{Gweon01b,Gweon01,Gweon02,Denlinger99}  A large
anomalous exponent $\theta = 0.9$ and differing values for the velocities
$u_\sigma$ and $u_\rho$ are required to reasonably account for the data.     According to the LL theory of the
spectral weight,\cite{Meden92}
    a
large
$\theta$  leads to   edges of holon and spinon  excitations that crosses the Fermi
edge with different amplitudes as shown in
Fig.~\ref{Lithium}-b. The LL theory used
in Fig. 13-b is for temperature $T=0$ and has $u_\sigma/u_\rho = 1/5$ as
in the analysis of Ref.~\cite{Denlinger99}.  Subsequent analysis \cite{Gweon02} of the data of
Fig. 13-a using a finite temperature theory \cite{Orgad01,Orgad01b} leads to a value of
 $u_\sigma/u_\rho $ = 1/2.  A large value of
$\theta $ for a system like Li$_{0.9}$Mo$_6$O$_{17}$ with incommensurate
band-filling  would   indicate  that
long-range Coulomb interactions  play  an important role in such a system.
According to the results of
\S~\ref{Fliquid}, a large
$\theta$ will also yield    strong electronic confinement along the chains.
The impact on other physical
properties should also be observable as for example in the poor
conductivity and the absence of a plasma
edge in  the transverse directions.\cite{Degiorgi88}

\section{Conclusion}
 A large part of the phenomenology shown by quasi-one-dimensional conductors cannot be understood in the
traditional framework of solid state physics. Probably for no other crystals do we have to reckon with concepts
 provided by the now well understood physics of interacting electrons in one dimension.   The
existence of long-range order at finite temperature in many of these  systems indicates that the link  
between one  and higher-dimensional physics is also essential to their understanding.

As we have seen in this review some progess has been achieved in that direction but there are also 
several basic questions left to answer. Among them, let us mention how Fermi liquid quasi-particles are 
appearing in the normal phase for  systems like the Bechgaard salts (or their sulfur analogs at high pressure) ? 
A clarification of this issue  would certainly represent  a significative advance in the comprehension
of these fascinating low-dimensional solids.

\subsubsection{Acknowledgements}
The author thanks D. J\'erome, L.G.
Caron, R. Duprat, N. Dupuis, J. P. Pouget, A.-M. Tremblay and R. Wortis  for numerous  discussions on several
aspects of this work; J. W. Allen and G.-H. Gweon  for several remarks and correspondance about ARPES results on
molybdenum bronzes.  This work is supported by  the Natural Sciences and Engineering Research Council of Canada
(NSERC), le Fonds pour la Formation de Chercheurs et l'Aide \`a la Recherche du Gouvernement du Qu\'ebec (FCAR),
the `superconductivity program' of the
Institut Canadien de Recherches Avanc\'ees (CIAR).
\bibliography{articles,Note}

\begin{thebibliography}{100}
\bibitem{LRO}
Strickly speaking, long-range order at finite temperature is impossible in two
  dimensions for an order parameter with more than one component. However, a
  finite coupling in the third direction is present which enables to obtain a
  finite $T_c$.
\end{thebibliography}
   \bibliographystyle{prsty}
\end{document}